\documentclass[a4paper]{aa}

\usepackage{xspace}
\usepackage{graphicx}
\usepackage{color}
\usepackage{txfonts}
\usepackage{natbib}
\usepackage{lscape}
\usepackage{multicol} 

\newcommand{\kms}{\, {\rm km\, s}^{-1}}

\newcommand{\eg}{{\it e.g.}\xspace}
\newcommand{\ie}{{\it i.e.}\xspace}

\def\RF{{\tt RingFinder}\xspace}
\def\slfit{{\tt sl\_fit}\xspace}
\def\slmock{{\tt sl\_mock}\xspace}
\def\galfit{{\tt galfit}\xspace}

\def\hst{{\it HST}{ }}
\def\cfht{{\it CFHT}{ }}

\def\Reff{R_{\rm eff}}
\def\REin{R_{\rm Ein}}

\begin{document}

\title{Extensive light profile fitting of galaxy-scale strong lenses.}
\subtitle{Towards an automated lens detection method}

\author{ F.~Brault\inst{1}, R.~Gavazzi\inst{1}}

\institute{
 Institut d'Astrophysique de Paris, UMR7095 CNRS \& Universit\'e Pierre et Marie Curie, 98bis Bd Arago, F-75014, Paris, France
}

\date{\today}


\abstract
{}
{
We investigate the merits of a massive forward modeling of ground-based optical imaging  as a diagnostic for the strong lensing nature of Early-Type Galaxies, in the light of which blurred and faint Einstein rings can hide.
}
{
We simulate several thousand mock strong lenses under ground- and space-based conditions as arising from the deflection of an exponential disk by a foreground de Vaucouleurs light profile whose lensing potential is described by a Singular Isothermal Ellipsoid. We then fit for the lensed light distribution with \slfit after having subtracted the foreground light emission off (ideal case) and also after having fitted the deflector's light with \galfit. By setting thresholds in the output parameter space, we can decide the lens/not-a-lens status of each system. We finally apply our strategy to a sample of 517 lens candidates present in the CFHTLS data to test the consistency of our selection approach.}
{
The efficiency of the fast modeling method at recovering the main lens parameters like Einstein radius, total magnification or total lensed flux, is quite comparable under \cfht and \hst conditions when the deflector is perfectly subtracted off (only possible in simulations), fostering a sharp distinction between the good and the bad candidates. Conversely, for a more realistic subtraction, a substantial fraction of the lensed light is absorbed into the deflector's model, which biases the subsequent fitting of the rings and then disturbs the selection process. We quantify completeness and purity of the lens finding method in both cases. 
}
{
This suggests that the main limitation currently resides in the subtraction of the foreground light. Provided further enhancement of the latter, the direct forward modeling of large numbers of galaxy-galaxy strong lenses thus appears tractable and could constitute a competitive lens finder in the next generation of wide-field imaging surveys.
 }

   
\authorrunning{Brault \& Gavazzi}
\titlerunning{Automated lens modeling}

\maketitle


\section{Introduction}

Strong gravitational lensing can be used to great effect to measure the mass profiles of early-type
galaxies, both in the nearby universe and at cosmological distances
\citep[e.g.\ ][]{T+K02a,T+K02b,RKK03,T+K04,Koo++06,J+K07,Gav++07,Tre10,Aug++10,Lag++10,Son++12,Bol++12,Dye++14,ORF13}.
Until recently, this approach was severely limited by the
small size of the samples of known strong gravitational lenses. 
With the advent of deep, wide-field imaging surveys we have now
entered a new era that enables the use of sizable samples of strong
lensing events as precision diagnostics of the physical properties of
the distant Universe.

The most abundant configurations of strong lensing events observed to date
involve a foreground massive galaxy and a background somewhat fainter source that is well
aligned with the former (so-called galaxy-galaxy strong lensing).
Even though other recent approaches based on spectroscopic surveys \citep{Bol++04,Bol++06,Aug++10,Bro++12}
or imaging data at sub-mm and mm wavelengths \citep{Neg++10,War++13,Vie++13} have been very successful,
there is strong motivation to develop techniques to identify
many galaxy-galaxy strong lenses purely in very abundant optical imaging data.
\citet{MBS05} forecast more than 10 such systems per square degree at \hst-like depth and resolution.
Still, ground-based imaging is a potentially promising avenue thanks to the ready
availability of hundreds of square degrees in multiple
bands with sub arc second image quality like the Canada France Hawaii Legacy Survey (CFHTLS).

This argument motivated the Strong Lensing
Legacy Survey (SL2S) which comprised, in addition to a search for group and cluster
scale (Einstein Radius $\REin\gtrsim 3\arcsec$) lenses
\citep{Cab++07,Mor++12}, a galaxy-scale lens search based on the \RF technique
\citep[][hereafter G14]{Gav++14}.
The main goal of this latter work was to use newly found lenses at redshift $z\sim 0.6$
to study the formation and evolution of massive galaxies through a joint strong lensing and stellar-dynamical method \citep{Ruf++11,Gav++12,PaperIII,PaperIV,PaperV}.
The main ingredient behind the \RF approach of finding galaxy-galaxy strong lenses resides in a simple
and fast difference imaging technique in two bands.
Although a fair level of automation and reproducibility was reached by \RF, it would be a step forward to gain further insight on the lens/not-a-lens nature of a lens candidate in order to achieve purer samples: that would indeed reduce the human intervention currently needed to visually classify the systems, as well as the follow-up operations that could get quite expensive for the very large surveys to come.

To this end, we propose a new lens finder that would encode the lens equation through the massive modeling of all candidates. So far, fine lens models were performed after the selection process as a way to (in)validate the lens nature of the candidates and give accurate estimates of the lens parameters if appropriate. In the case of this new finder, it would consist of implementing a fast modeling mode that would precede the decision phase on the candidates status, thus requiring to make  a tradeoff between speed and robustness. Other improvements can be made at the level of the deflector's light subtraction \citep[see e.g.,][]{Jos++14} that might relax the \RF need of two bands and make the selection more complete regarding lenses where the source is not blue. Therefore, a promising strategy would be to quickly fit to single-band imaging data a lens model featuring a simple background source and a simple foreground deflector with a small number of degrees of freedom. Some ingredients of such a method have been implemented by \citet{Mar++09} on archival \hst data, as part of the HAGGLES effort \citep[see also,][]{NMT09}. The assumption of surface brightness conservation made in HAGGLES is sensible in the latter case of \hst image and allows for fast modeling shortcuts. This is however not possible for ground-based imaging with seeing of size comparable to the size of lensed features. One has therefore to fully implement the convolution by a large Point Spread Function (PSF) and carefully disentangle the light coming from the foreground object and that of the background lensed source before any attempt to quickly and robustly model the observed light distribution.

In this paper we attempt to immerse a versatile lens modeling code, \slfit \citep[][]{Gav++07,Gav++08,Gav++11,Gav++12}, into a pipeline that will rapidly fit the light distribution around lens candidates with a simple lens model featuring a foreground Early-Type Galaxy (ETG), a background source, and a Singular Isothermal Ellipsoid (SIE) mass distribution. 
We first used simulated data to test the speed and reliability of our model optimization techniques, to finally adjust the tradeoff between  a fast agressive modeling and a good exploration of the parameter space. The ingredients entering the simulations as well as the modeling assumptions are presented in Sect.~\ref{sec:model} while we report on the precision and speed in the recovery of key model parameters in Sect.~\ref{sec:resmock}.
The next step is to find a set of quantities provided by the output best fit model that will optimally allow to separate the population of simulated lenses from the population of simulated non-lenses. This is the object of Sect.~\ref{sec:resmock:robot}. Finally, in Sect.~\ref{sec:resreal}, we perform the same modeling approach on the homogeneous sample of SL2S lenses (G14) enriched with additional preselected systems also found in the CFHTLS data, in order to see how much we can improve over the existing selection, both in terms of completeness and purity.
We summarize our main results and present our conclusions in Sect.~\ref{sec:conc}.

\section{Simulating and Modeling strong lenses}\label{sec:model}

\subsection{Physical assumptions}\label{sec:model:assump}
For the simulations as well as for the subsequent modeling part, we make the same assumptions as the simulations of G14. We briefly recap the features that are relevant for this work.

The systems involve a foreground deflector whose light distribution is described by an elliptical de Vaucouleurs profile which reads:
\begin{equation}
I(r) = I_0 \exp\left[ -k_4 \left(\frac{r}{\Reff}\right)^{1/4} \right],
\end{equation}
with $k_4 \simeq 7.669$ and $\Reff$ is the effective  or half-light radius and $I_0$ the central surface brightness. To capture elliptical symmetry, we also need to introduce an axial ratio $q$ and a position angle $\phi$. The deflector will always be centered at the origin of the coordinate system. 

The total mass profile attached to the deflector is described by a Singular Isothermal Ellipsoid. In the simulation, the ellipticity and orientation of the light and mass are matched although this is partly relaxed in the modeling (see below). The strength of the mass distribution, that is its Einstein radius, is given by the velocity dispersion of the isothermal sphere:
\begin{eqnarray}
\REin & = &4\pi \left( \frac{\sigma}{c}\right)^2 \frac{D_{\rm ls}}{D_{\rm s}} \:{\rm rd} \\
      & = &\left( \frac{\sigma}{186.2 \kms}\right)^2 \frac{D_{\rm ls}}{D_{\rm s}} \:{\rm arcsec}\;,
\end{eqnarray}
where $D_{\rm ls}/D_{\rm s}$ is the ratio of angular distance between the lens and the source, and between the observer and the source.
The luminous content and the mass are tightly coupled by scaling relations in the simulation as described in G14.

A background source with an elliptical exponential light distribution is also placed behind the deflector.
The population of lenses and sources are taken from the CFHT-LS catalog of $i<22$ early-type galaxies and the COSMOS catalog of faint $i<25$ objects, respectively \citep{Cap++07b,Ilb++09}. Hence, the complex covariance of magnitude, effective radius, and redshifts is naturally captured by these real catalogs. 

The simulations and the model fitting are consistently performed with \slmock~and \slfit that were introduced in G14 and \citep{Gav++07,Gav++11,Gav++12}, respectively. Those two codes share the same image rendering and lensing engines.

\subsection{Observational conditions and preparation of the data}\label{sec:model:cond}
In order to quantify how well one can recover lens model parameters, we want our simulated lenses to mimic two typical observing conditions for a given input light distribution.
We thus focus on a ground-based survey of depth and image quality given to the CFHT-LS Wide (cf \S~\ref{ssec:cfhtls}).  Seeing conditions and exposure times were randomly taken from real CFHTLS conditions.  A library of 50 PSF models spanning a broad realistic range of seeing conditions for this survey is built.  The input PSFs are directly inferred by sub-pixel centering of  the stars in the CFHTLS data. Such PSF images are constructed at the natural $0\farcs186$ Megacam pixel scale.
We simulate observations in the three $g, r, i$ optical channels of CFHT Megacam and mock deflectors are produced down to a magnitude $i\lesssim 22$. A noise model  that captures the contributions of the photon noise of sky and the objects along with readout noise is applied and stored in variance maps. Hence the noise is non stationary but is very well described by gaussian statistics of zero mean and dispersion $\sigma_{ij}$ for a given pixel $ij$.

In addition to those ground-based observing conditions, we simulate the same objects under space-based observations as it was obtained with the WFPC2 camera and F606W filter of the \hst during the follow-up campaign of SL2S lenses \citep{Gav++12,PaperIII}.  \hst observations are 1400 sec long and resemble the planned Euclid depth with a slightly redder filter and slightly broader PSF \citep{Euclid}.
In that case, images are sampled at a $0\farcs1$ pixel scale. We used PSF images calculated with {\tt TinyTim} \citep{kri++11}. Oversampling by a factor 4 was used in the \hst simulations, and the {\tt TinyTim} PSF were sampled accordingly.
However, in order to save computing time in the modeling phase, we did not refine the sampling of the light distribution predicted by a given model. We tested that this agressive image rendering does not lead to significant modeling errors. Therefore unless otherwise stated we will not apply the same $\times 4$ oversampling of the model and the PSF.

\begin{figure}[htb]
  \centering
  \includegraphics[width=\columnwidth]{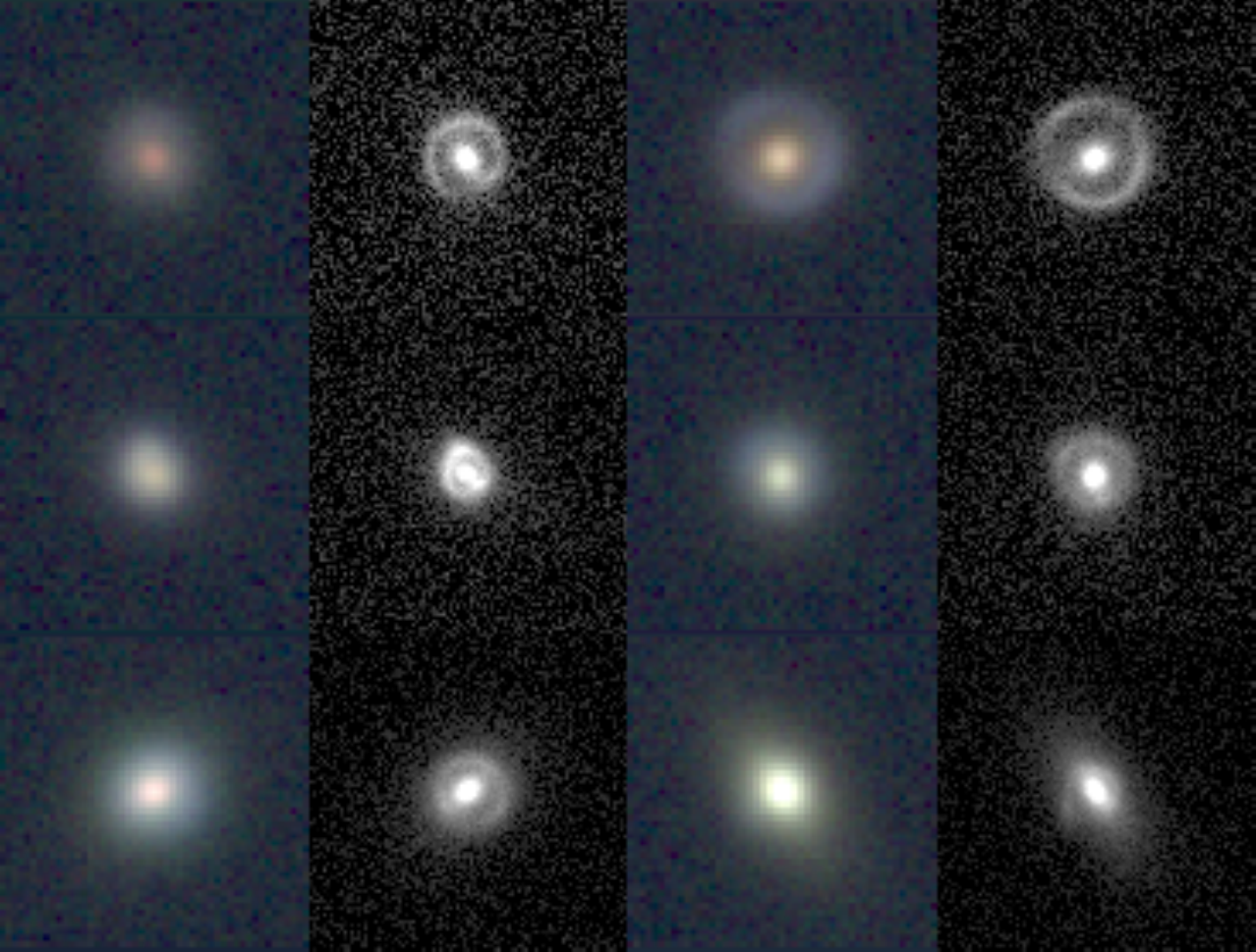}
  \caption{Mosaic of 6 simulated gravitational lenses under CFHT ($gri$-color, left) and HST/WFPC2 (gray-scale, right) conditions. Each panel is $\sim9\farcs2$ on a side.}\label{fig:SimIllust}
\end{figure}

The mock observations thus consist in 3 (CFHT) + 1 (WFPC2) noisy input images, along with 3+1 input variance map images, and 3+1 input PSF images. Simulated postage stamps are $9\farcs2$ on a side but we focus on a smaller fitting window that is $7\farcs7$ wide for the subsequent modeling phase. For illustration, we show in Fig.~\ref{fig:SimIllust} a few simulated systems under the ground and space conditions. 

\subsection{Subtraction of foreground deflector}\label{sec:model:defsub}
In addition, we are focusing here on the modeling of the lensed features that are most of the time well buried into the peak of light emitted by the foreground deflector. It is therefore necessary to subtract it off before undertaking a precise modeling of the lensed background emission. Thanks to our simulations, we can quantify the impact of imperfect subtraction on the modeling itself. Of course, the photon noise supplied by deflector's emission is already taken into account but we want to check whether the modeling and, in fine, the lens / not-a-lens decision diagnostic is affected. \citet{Mar++07} already found that, at least for high-resolution imaging, either from Adaptive Optics or from \hst~imaging, the Einstein radius is well recovered but that the less constrained parameters like source size or shape can be biased. With our simulations, we want to assert that this is also true for the fast modeling of ground-based data where it is even more difficult to disentangle the foreground and the background emissions. We therefore consider two kinds of deflector subtraction methods:
\begin{itemize}
\item an ideal reference case in which we have a perfect knowledge of the foreground emission. The latter is therefore trivially subtracted. This leaves us with lensed patterns (arcs, rings) that are just affected by noise of known statistics.
\item We also consider \galfit~\citep{Pen++02,Pen++10b} to subtract off a de Vaucouleurs light profile by trying to mask out the most prominent lensed features in an automated (hence not perfect) way. For ground-based data, we perform a 4-$\sigma$ clipping of the discrepant pixels in order to down-weight the effect of the lensed features on the fitting of the foreground emission. Moreover, since we are simulating observations in the three $gri$ channels, we perform a 4-step fitting, starting from the reddest $i$ band image (in which the deflector light dominates, on average). This model of the deflector is then used as a starting point for the fit in the $r$ band, which is then used as a starting point for the fit in the $g$ band. Again, we perform a 4-$\sigma$ clipping of discrepant pixels that makes a 1-0 mask that is then used for another fitting of foreground light emission with \galfit in $g$ band.
The same masking process will be applied when subtracting the deflector from observed CFHTLS candidates in Sect.~\ref{sec:resreal} using 4 bands ($griz$).
For space-based data, we remind that the simulations are carried out in only one filter. Nonetheless, the particular case of space conditions with a \galfit subtraction is treated with a finer sampling of the model and the PSF (see Sect. \ref{sec:model:cond}) to give a sense of the increased execution time. As explained below, this is the only situation in which the finer $\times 4$ sampling was somewhat needed.
\end{itemize}

\subsection{Bayesian inference}\label{sec:model:infer}

We now develop the methodology that we adopted to fit for the lens modeling parameters with the \slfit~tool in a Bayesian framework.

\subsubsection{Definitions}
The vector of observations, which is the set of pixel values in a small cutout region around a lens candidate, is $\vec{d}$. Pixel values are assumed to be devoid of light coming from the foreground deflector as discussed above.

The posterior probability distribution of the model parameters $\vec{\theta}$ given the observations $\vec{d}$ is given by the Bayes theorem:
\begin{equation}\label{eq:bayes}
\mathcal{P}(\vec{\theta} \vert \vec{d} ) = \frac{ \mathcal{L}( \vec{d} \vert \vec{\theta}) \times p(\vec{\theta})}{\mathcal{E}(\vec{d})}\,,
\end{equation}
with the usual definition of the likelihood $\mathcal{L}( \vec{d} \vert \vec{\theta})$, the prior $p(\vec{\theta})$ and evidence $\mathcal{E}(\vec{d})$, which reduces to a normalization constant here since we are testing just a family a models.

The likelihood can be built with the noise properties we have presented above. If we consider an image of size $n\times n$ with pixel values stored in a model vector $\vec{m}(\vec{\theta})$, we can write:
\begin{equation}
\mathcal{L}( \vec{d} \vert \vec{\theta} ) =  (2 \pi)^{-n^2/2} \prod_{i=1}^{n^2} \sigma_i^{-1} \exp\left[ -\frac{(d_i-m_i(\vec{\theta}))^2}{2\sigma_i^2}\right] \propto {\rm e}^{-\chi^2/2}\;.
\end{equation}

The assumptions developed in Sect.~\ref{sec:model:assump} allow to predict the model image $\vec{m}(\vec{\theta})$ and specify the parameters contained in $\vec{\theta}$. In Table~\ref{tab:params}, we list them as well as the two kinds of priors we assume for them. For the situation of a \galfit subtraction of the deflector, the central values of the priors on the lens potential axis ratio $q_{L}$ and orientation $\phi_{L}$ are taken from the best fit value of the \galfit optimization. We also use the center of the light profile fitted by \galfit as the center of the potential $(x_{c},y_{c})$ for the \slfit modeling, assuming that mass follows light. The case of an ideal subtraction has been run afterwards so that we could keep the same values of $q_{L}$ and $\phi_L$ for more consistency between the two subtraction modes; yet $(x_{c},y_{c})$ are obviously set to $(0,0)$ in this case as for the simulations. 

\begin{table}
\centering
\caption{\label{tab:params} Definition of model parameters and associated priors}
\begin{tabular}{lc}\hline\hline
parameter & Prior \\\hline
\multicolumn{2}{c}{Lens potential} \\
axis ratio $q_p$  &  $\mathcal{N}(q_{\rm L}\tablefootmark{a}, 0.2) \times \mathcal{U}(0.15,0.999) $ \\
orientation $\phi_p$  &  $ \mathcal{N}(\phi_{\rm L}\tablefootmark{a},30^\circ) \times \mathcal{U}(-90^\circ,90^\circ) $  \\
Einstein radius $\REin$ &   ${\rm log}\mathcal{N}(0\farcs17,0\farcs4)\times \mathcal{U}(0\farcs,5\farcs)$ \\
\multicolumn{2}{c}{Lensed Source} \\
center $x_c$   &  $\mathcal{N}(0.,0.4) \times \mathcal{U}(-1\farcs6,1\farcs6)$ \\
center $y_c$   &  $\mathcal{N}(0.,0.4) \times \mathcal{U}(-1\farcs6,1\farcs6)$ \\
axis ratio $q_s$ &  $\mathcal{N}(0.52,0.2)\times \mathcal{U}(0,1) $   \\
orientation $\phi_s$  & $\mathcal{U}(-90^\circ,90^\circ)$ \\
magnitude $m_s$ &   $\mathcal{N}(24.5,1.5)$ \\
size ${\Reff}_s/{\rm arcsec}$ &  ${\rm log}\mathcal{N}(-0.5,0.3) \times \mathcal{U}(0.01,2.) $ \\\hline
\end{tabular}
\tablefoot{$\mathcal{U}(a,b)$ stands for a uniform distribution between $a$ and $b$.
$\mathcal{N}(\mu,\sigma)$ stands for a normal distribution centered on $\mu$ and dispersion $\sigma$.
${\rm log}\mathcal{N}(m,s)$ stands for a log-normal distribution centered on $\log_{10}(x)=m$ and dispersion $s$.
\tablefoottext{a}{$q_L$ and $\phi_L$ are taken from the best fit value of the \galfit optimization.} 
}
\end{table}

\subsubsection{Optimization scheme}\label{sssec:optim}
Since we want to achieve a careful exploration of the degeneracies in the space of free parameters, we consider a full sampling of the posterior probability distribution with Monte Carlo Markov Chain (MCMC) techniques. The practical implementation in \slfit~is the Metropolis-Hastings algorithm. The convergence is accelerated by a burn-in phase (so-called "simulated annealing") in which, starting from a pure sampling of the prior, we increasingly give more weight to the likelihood. Defining a pseudo-temperature $T_n$ that decreases with the iteration number $n$, this reads:
\begin{equation}
 \mathcal{P}_n(\vec{\theta}) = \mathcal{L}(\vec{\theta})^{1/T_n} p(\vec{\theta}).
\end{equation}

After the cooling phase that guides the chain towards the global maximum a posteriori, we can correctly sample the posterior near its maximum. We empirically found that using a total number of iterations $N_{\rm tot}=2700$ was a good tradeoff between speed and robustness. The last 300 iterations really sample the posterior and the final 100 of them are used for the subsequent statistical analysis of the chain (best fit parameters and error)\footnote{The median and the 16th and 84th percentiles on the marginal parameter posterior distribution are used to define central values and confidence intervals.}. Hence, our analysis is performed on the most converged part of the chain. 
In the perspective of massive modeling of many lens candidates, it does not appear so important to have an extremely accurate description of the confidence interval on each model parameter. The MCMC effort remains a robust way to sample complex degeneracies and, for instance, check whether the lensing hypothesis of a given system (that is, {\it "The probability that a background source is magnified by more than a certain value is greater than some other value"}) is consistent with the data.

Despite those benefits, the cost of a full sampling is large in terms of CPU time. Calculations were performed in parallel with modern $2\,$GHz, 12-core {\tt x86\_64} servers. As we shall see below, it takes of order $10$ seconds to run a MCMC of length $N_{\rm tot}=2700$ iterations on CFHT $g$ band cutout images.

In order to reduce further the execution time in view of a massive modeling sequence for automatic lens detection, we explored the possibility to renounce the benefits of a Monte-Carlo sampling of the posterior PDF and used the faster simplex optimization scheme based on the {\tt minuit} library\footnote{\url{http://seal.web.cern.ch/seal/work-packages/mathlibs/minuit/home.html}}. Gradient methods were also envisioned but turned out not being robust enough in the presence of strong degeneracies between model parameters and multiple local minima, especially for realistic \galfit subtractions of the deflector. An option of {\tt minuit} allows the algorithm to refine its estimate of the parameter covariance matrix at run time and to account for non-symmetric errors on best-fit parameters. This practicality generally led to many more computationally expensive likelihood estimations, to such a point that it is not competitive with and less robust than our fiducial MCMC method, even in the case of an ideal subtraction. Consequently, we do not retain this fast minimization option in its current implementation in our code as it is not robust enough and produces too many catastrophic failures (especially for a realistic \galfit subtraction of the deflector). Nonetheless, the reduced CPU time consumption is a good incentive for improving the parameterization of the  problem (to reduce degeneracies), finding better convergence criteria, parallel optimizations, simulated annealing in the {\tt minuit} phase, etc.

\section{The recovery of key lens parameters on mock data}\label{sec:resmock}
In this section, we assess the recovery by \slfit of the most relevant quantities that could help deciding whether a foreground galaxy acts as a strong lens on a background source or not. We consider both the CFHTLS-like $g$ band\footnote{For the modeling of the lensed features, it is desirable to maximize the relative flux contribution of the blue background source over the redder foreground deflector.} and \hst WFPC2 simulated images. The most important quantities are the Einstein radius $\REin$, the total magnification $\mu_{\rm tot}$, and the total image plane $g$ band magnitude $m_{\rm img}$ of the lensed light.

We distinguish the ideal situation of a perfectly subtracted foreground deflector in Sect.~\ref{sec:resmock:ideal} and the more realistic case of a preliminary subtraction performed with \galfit  in Sect.~\ref{sec:resmock:galfit} to show how much the complex crosstalk between foreground and background light (especially for the bright highly magnified sources) has to be treated with caution.

Our deep parent sample of COSMOS sources provides many lensed objects that are too faint to be detectable under CFHTLS conditions or to provide sufficient information for the modeling. Systems having a total magnified flux such that $m_{\rm img,0} > 24.5$ are therefore not taken into account in the rest of the analysis. 

\begin{figure}[htb]
  \centering
  \includegraphics[width=4.45cm]{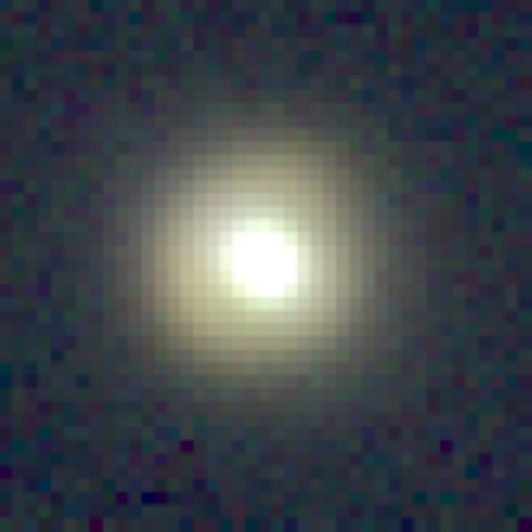}
  \includegraphics[width=4.45cm]{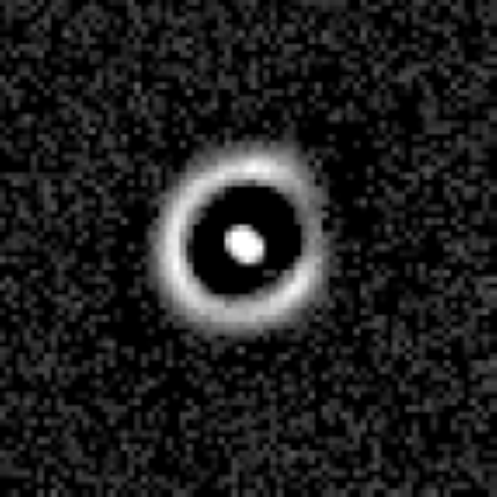}
  \includegraphics[width=4.45cm]{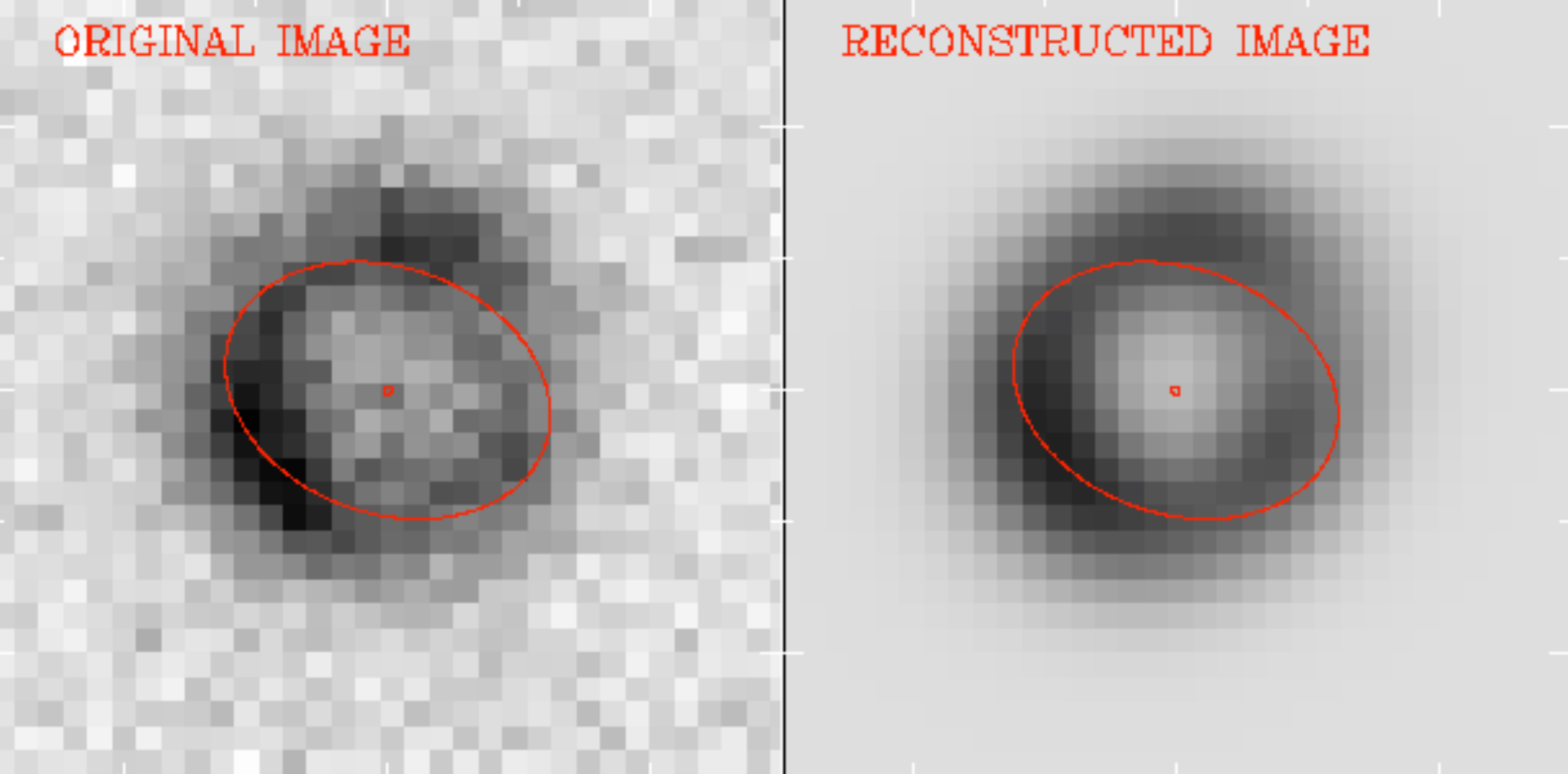}
  \includegraphics[width=4.45cm]{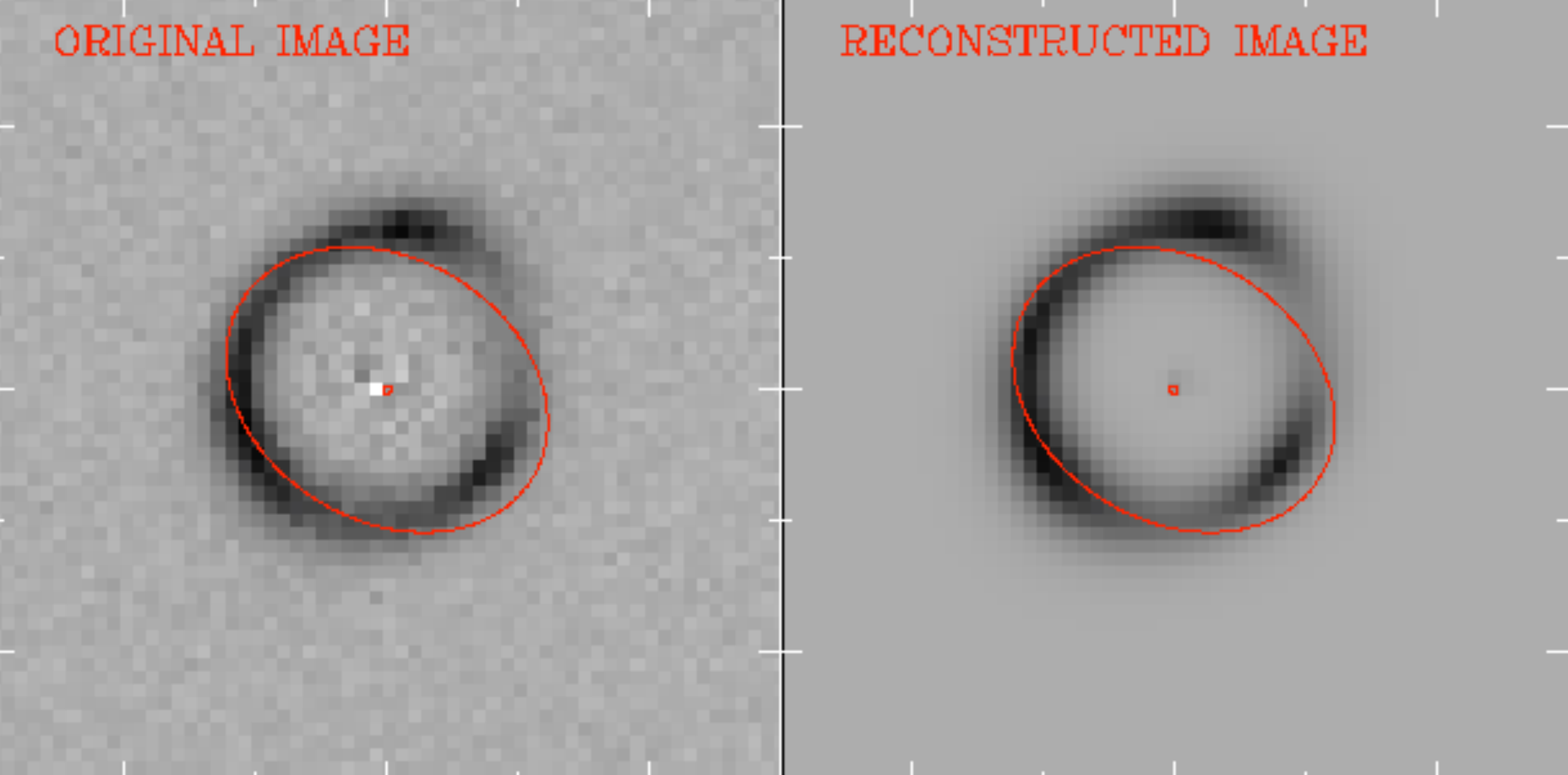}
  \includegraphics[width=4.45cm]{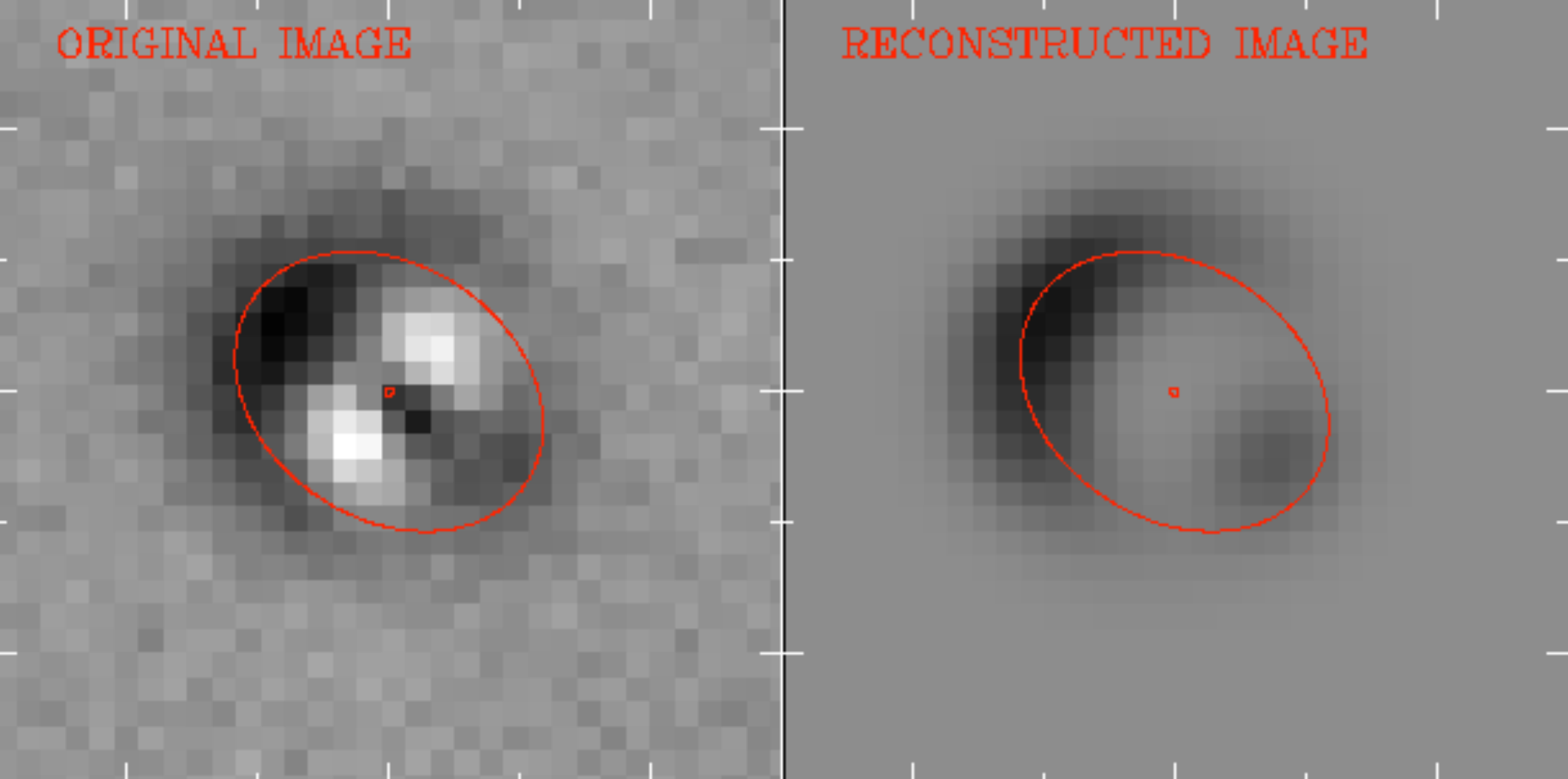}
  \includegraphics[width=4.45cm]{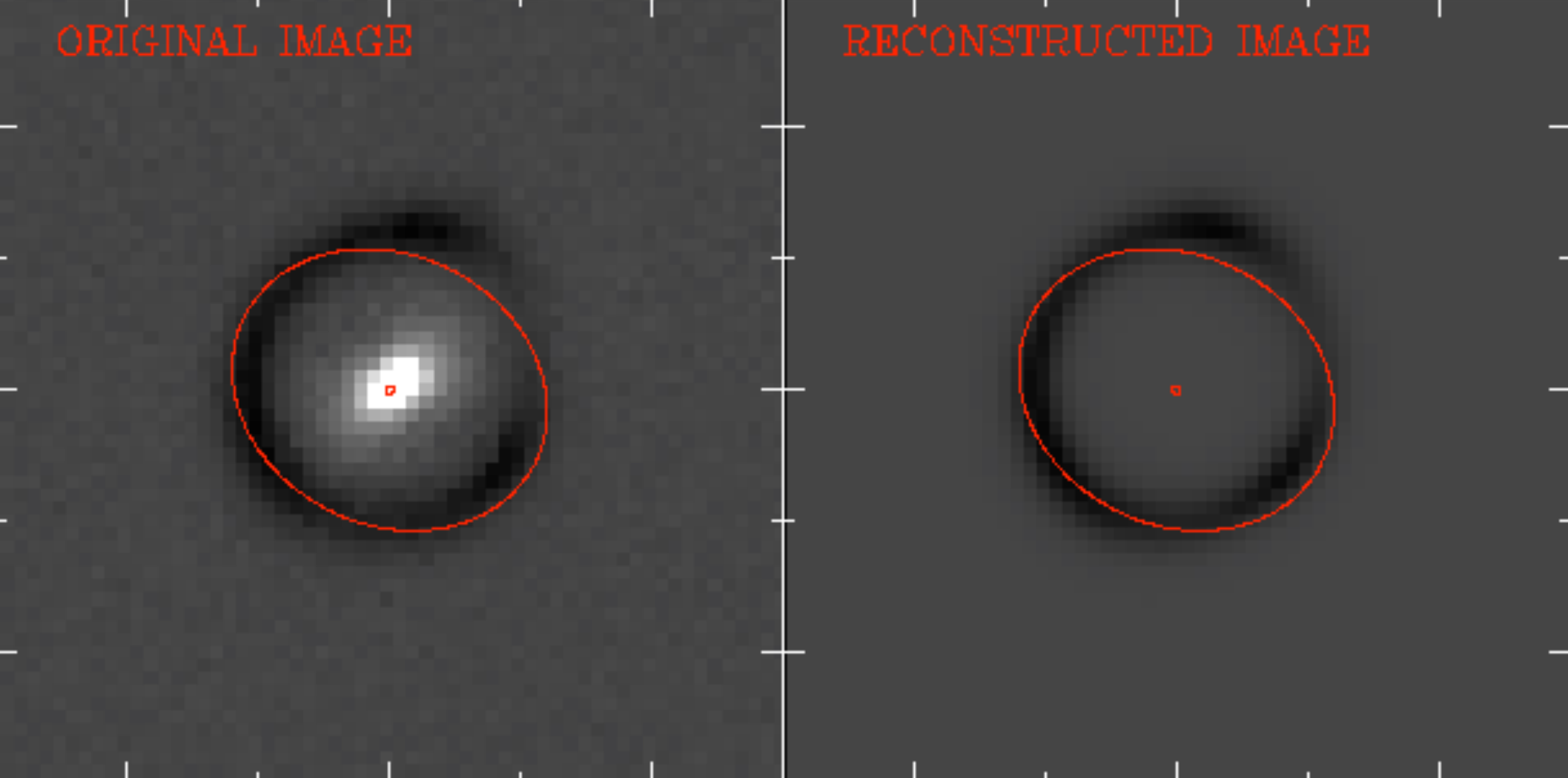}
      \caption{Comparative modeling of a simulated ring in the 4 cases considered in Sect.~\ref{sec:resmock}. The top panels show the mock observations of the lens system under CFHT (gri-color, \textit{left}) and HST/WFPC2 (gray-scale, \textit{right}) conditions. The other panels show images where the deflector has been subtracted ideally \textit{(middle panels)} or using \galfit \textit{(bottom panels)}, along with the lensed features reconstructed by the model and the critical lines \textit{(in red)}. The CFHT simulations are modelled in the $g$ band. Each panel is $7\farcs7$ on a side. Though the original lensing configuration is not perfectly recovered in the case of a \galfit subtraction of the deflector from the mock ground-based observation, we see that the brightest parts of the ring are still identified, allowing a robust estimation of $\REin$. In this example we find $(\REin-{\REin}_{0})\sim 0\farcs017$ (for an initial value ${\REin}_{0}\sim 1\farcs129$), and the reduced $\chi^2 / \nu \sim 3.4$ per degree of freedom. }\label{fig:compmodel}
\end{figure}

Fig.~\ref{fig:compmodel} illustrates the results of such an automated lens modeling analysis of a simulated ring. Table~\ref{tab:resmock} gathers the main quantitative results. Since we want to check that the important lensing parameters are correctly recovered for the most valuable lensing configurations, we mostly consider large magnification systems by restricting the analysis of this section to configurations with a small impact parameter $\beta_0 \le 0\farcs2$. The numbers regarding the scatter and bias in the table could change somewhat for large radius less magnified lenses. We checked on a small number of available simulations that the bias is not significantly changed for such systems; obviously, the scatter is slightly increased. Nevertheless we fully account for this increased dispersion of low magnification systems in the next section when we shall exploit \slfit as a lens finding tool.

Throughout this article, the subscript $0$ refers to input simulation quantities whereas quantities without this index refer to values actually recovered by \slfit.

\subsection{Comparing ground and space performance with ideal subtraction of the deflector}\label{sec:resmock:ideal}
\begin{figure}[htb]
  \centering
  \includegraphics[width=\columnwidth]{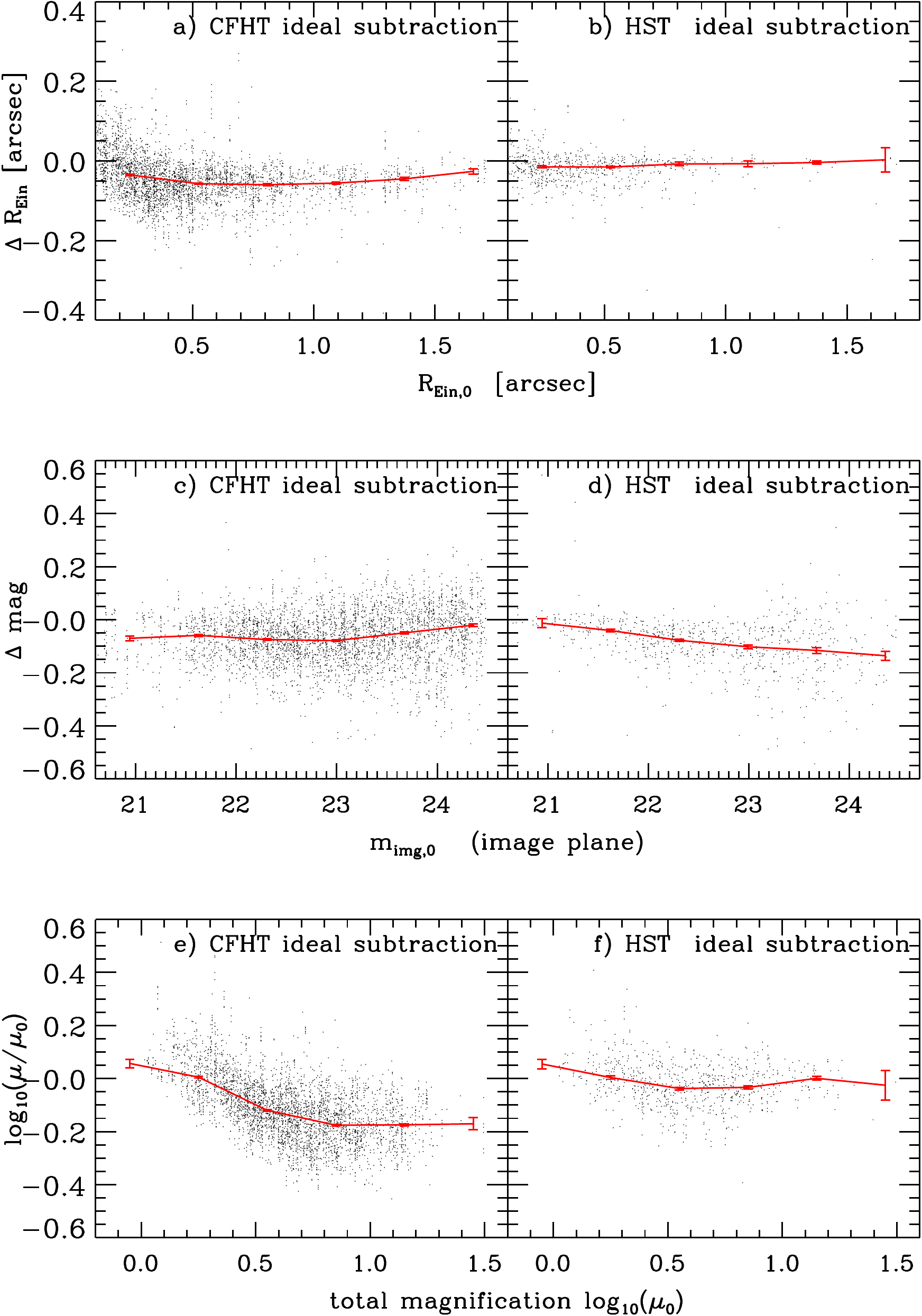}
  \caption{Recovery of Einstein Radius (top), total image plane lensed flux (middle) and magnification (bottom) for simulated lenses fitted with \slfit in fast modeling mode. We distinguish the case of ground-based CFHTLS-Wide-like $g$-band observations (left) and the case of \hst~WFPC2/F606W observations (right). In both cases, the deflector's light has been "perfectly" subtracted off before modeling the residual lensed features. In each panel, the red error bars show the binned median values for guidance.}\label{fig:SimIdeal}
\end{figure}
Fig.~\ref{fig:SimIdeal} shows the difference between recovered and input quantities for the Einstein Radius, the image plane total magnitude of the lensed light and the total magnification, respectively.

The Einstein radius is well recovered in the two cases, especially for the largest values which correspond to complete and bright arcs. We only observe a small bias of order $\langle \Delta \REin \rangle \simeq -0\farcs065$ from the ground based mock observations for values ${\REin}_{0} \gtrsim 0\farcs7$, and $\langle \Delta \REin \rangle \simeq -0\farcs017$ from the space counterparts. This is about a third and a sixth of a pixel respectively. At smaller radii (${\REin}_{0} \lesssim 0\farcs7$), we notice more scatter in the recovery of $\REin$ for CFHT than for \hst mock data : indeed, this population of objects contains an important fraction of faint, low magnification systems that are poorly resolved from the ground, leaving the model under-constrained.
 
We also observe a very small bias with a small scatter in the recovery of the total magnified flux of the lensed features, either from ground-based images or from space-based ones. Nevertheless, some outliers appear from HST simulations that did not exist for \cfht simulations : that might be due to the lower resolution of \cfht simulations that leads to a smoother parameter space for the walk of the Markov chain, the convergence of which is made easier.  After re-running the modeling procedure on these systems under the same conditions, most of them can be recovered with a value close to the initial one, meaning that the optimization had indeed fallen in a local extremum of the posterior during the first runs, hence illustrating the random character of the MCMC sampling. In the end, those outliers appear negligible, with a rate of a few thousandths. 
 
Despite the good estimation of $m_{\rm img}$, we can see a non negligible bias in the recovered total magnification $\mu_{\rm tot}$ from \cfht images for input magnifications above $\mu_0\sim 2.5$ as can be seen in the bottom left panel of Fig.~\ref{fig:SimIdeal}. This underestimation bias is not observed for \hst~simulations. We studied the possible origin of this bias and came to the conclusion that it could be due to a slightly too aggressive sampling of the \cfht model when performing a fast modeling.
Since $m_{\rm img}$ is well recovered and $m_{\rm img} = m_{\rm src} - 2.5 \log_{10} \mu_{\rm tot}$, this bias is reflected on the recovery of the intrinsic source flux, which is slightly overestimated for \cfht simulations and is good for \hst simulations. 
 Here we found $\mu\sim 0.7 \mu_{0} $, corresponding to $m_{\rm src} \sim m_{\rm src,0} - 0.365$.			

Concerning less magnified lensing configurations at large radii, we had a few such simulated systems at our disposal and checked that they mostly add little scatter in the recovery of $\REin$ and associated parameters without producing noticeable biases.
									
For ground-based simulations, the median time for the optimization to complete (\ie to reach 2700 iterations) is about $\sim 10$ seconds.
Only 1\% of the chains never converged, and some others took a very long time to complete. We quantify this "failure mode" with the fraction of optimizations that did not converge in less than 1 minute. This fraction is of order 9\% for \cfht simulations.
For \hst~simulations, those numbers rise to a median execution time of $\sim 18\,$s, and a failure rate of $\sim 16\%$ for a threshold value of  2 minutes.
The factor $\sim 2$ finer pixel scale of space data explains this increased time.
However, we also conclude that our calculations are not yet limited by the rendering of pixels (execution time would scale as $N_{\rm pixels}^2$) but rather by a combination of overheads (disk accesses, MCMC engine, etc) and a speed-up due to the better defined maximum of the likelihood surface for space data.

\subsection{The effect of imperfect \galfit subtraction of the deflector}\label{sec:resmock:galfit}
The previous section was optimistically assuming that the light from the foreground deflector could successfully
be removed without any implication on the lensed patterns. First of all, any attempt to disentangle foreground and background emissions is not an easy task for galaxies of a size comparable to the seeing disk and emitting at similar optical wavelengths. In Sect.~\ref{sec:model:defsub}, we proposed a method to mitigate the effect of the background emission when fitting the foreground deflector's light with \galfit and automated masks.

\begin{figure}[htb]
  \centering
    \includegraphics[width=\columnwidth]{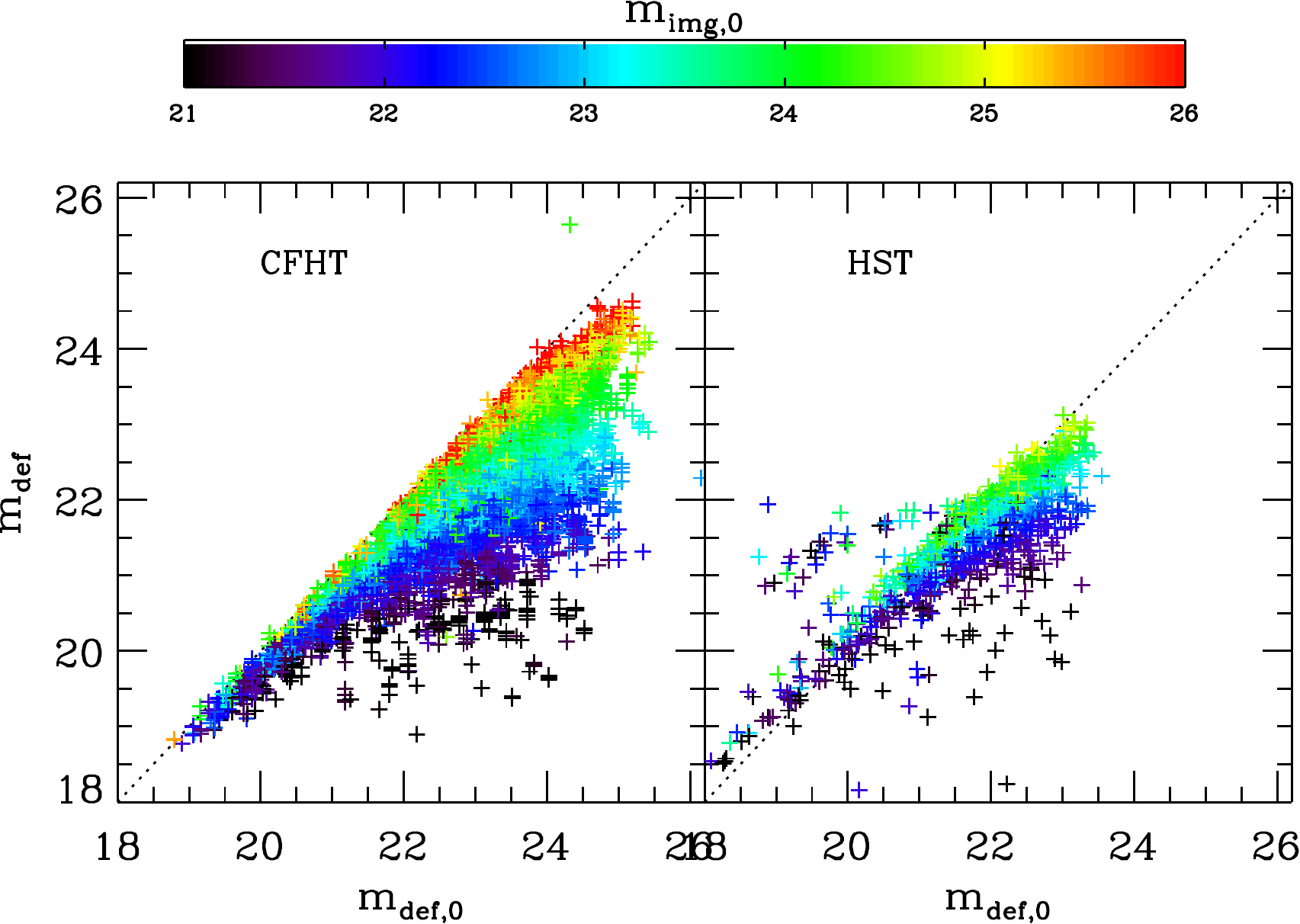}
  \caption{Recovery of the deflector's $g$ band flux when fitted with \galfit on CFHT ({\it left}) and \hst ({\it right}) simulations, colored according to the initial magnitude of the lensed light. We see some over-subtraction of the lensed background component for most cases.}\label{fig:DefGalfit}
\end{figure}
Fig.~\ref{fig:DefGalfit} shows the limits of the method. In the left panel, we see that the recovered deflector's flux is almost always overestimated for CFHT observations, with a stronger bias for brighter lensed features. This implies that some flux from this latter component has been captured by \galfit as belonging to the foreground emission. This reduction of the (large scale) flux of the arcs is, of course, not desirable and could potentially prevent a lens identification. The right panel concerns \hst simulations. We see that the nuisance is still present, though less pronounced at the faint end; we also clearly notice that the subtraction of the deflector light is less successful for higher values of  $\Delta m_{0} = m_{\rm img,0} - m_{\rm def,0}$ (where $m_{\rm def,0}$ is the deflector's magnitude),
especially at the bright end, leaving strong leftover emission in \slfit input image.
We checked that a more careful masking of the lensed features, similar to what can be done one-by-one on a small number of systems \citep[\eg][]{Bol++06,Mar++07}, alleviates the catastrophic (under)-subtraction of bright deflectors.

\begin{figure}[htb]
  \centering
   \includegraphics[width=\columnwidth]{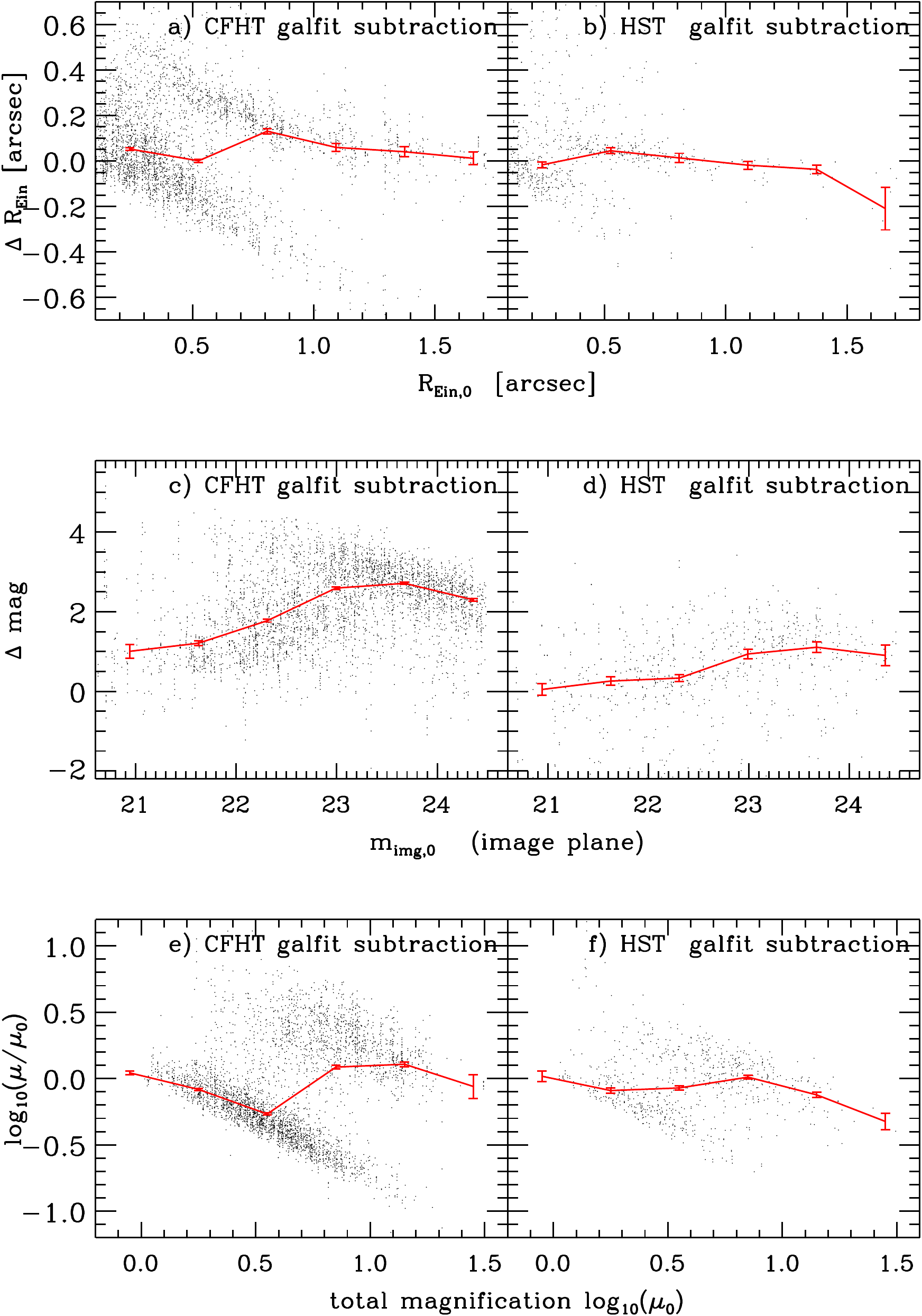}
  \caption{Recovery of Einstein Radius (top), total image plane lensed flux (middle) and magnification (bottom) for simulated lenses fitted with \slfit in fast modeling mode. We distinguish the case of ground-based CFHTLS-Wide-like $g$-band observations (left) and the case of \hst~WFPC2/F606W observations (right). In both cases, the deflector's light has been subtracted off with \galfit before modeling the residuals. In this case, over-subtracted background emission or under-subtraction or inaccurate modeling of the deflector produces larger errors and biases on fitted important parameters of the lens models.
  In each panel, the red error bars show the binned median values for guidance.}\label{fig:SimGalfit}
\end{figure}

The response of the modeling of the lensed features in the presence of those subtraction defects is complex.
We see in the top panels of Fig.~\ref{fig:SimGalfit} that the Einstein radius is biased in
a non trivial way for ground-based conditions when $\REin\lesssim 1\arcsec$.
It is less so for space, but still present for $\REin \lesssim 0\farcs7$.
Obviously, when the lensed features are far away from the deflector, the deflector's subtraction tends to the previously discussed case of an ideal subtraction.\\
At small radii ${\REin}_{0}\lesssim0\farcs3$, the simulated lenses are dominated by double systems with a main arc and a counter-image. From \cfht simulations, the subtraction generally contains little flux because \galfit has fitted most of their flux as part of the deflector flux. Due to this lack of constraints, \slfit is mainly driven by the priors as it would be for very faint lenses. \hst resolution helps identifying the gravitational arcs even when they are faint or very entangled with the deflector : in both cases, \galfit absorbs the counter-image, so that the subtraction shows a singly-imaged configuration, leading to an underestimation of the Einstein radius with a characteristic lower bound  $\REin \lesssim {\REin}_0/2$. \\
At intermediate radii, ($0\farcs3\lesssim {\REin}_0\lesssim1\farcs$ for CFHT, $0\farcs2\lesssim {\REin}_0\lesssim0\farcs6$ for \hst), our population of mock lenses already provides a majority of rings or quasi-rings. Concerning \cfht simulations, two net branches show up. For objects where \galfit still borrows a big fraction of the lensed flux, the larger Einstein radius makes some part the arc (not always the brighter part) stand out from the rest of the system, ending up with a singly-imaged lens configuration layout: the Einstein radius is thus underestimated and those lenses make the lower branch of the figure. The upper branch is due to situations in which \galfit subtraction leaves a symmetrical residual looking like an Einstein ring encompassing a hole of negative values. The typical radius is limited by the PSF FWHM $IQ$. This forces the lens model to predict a ring-like system with $\REin \simeq \max({\REin}_0, IQ)$ of greater magnification, though with underestimated overall flux (bottom and middle panel). In \hst simulations, at these radii, the resolution has already disentangled the lensed features from the deflector : the latter is then easily subtracted, taking just little flux from the ring, barely affecting its position or width. Consequently, the lower branch observed from \cfht simulations almost disappears and the upper branch is barely reduced.\\
 Finally, the largest Einstein radii (${\REin}_{0}\gtrsim1\farcs$ for CFHT and, ${\REin}_{0}\gtrsim 0\farcs7$ for \hst) are quite well recovered as the rings are completely disentangled from the deflector in both \hst and \cfht simulations. The few systems in the \hst simulations that have large $|\Delta \REin|=|\REin-{\REin}_{0}|$ are due to the aforementioned failures of the foreground subtraction for some bright objects.
\begin{table}
\centering
\caption{\label{tab:resmock} Modeling results on the simulated lenses}
\begin{tabular}{lcccc}\hline\hline
  & \multicolumn{2}{c}{CFHT $g$-band} & \multicolumn{2}{c}{HST WFPC2}  \\
  &  ideal  & \galfit   & ideal  & \galfit   \\
 &   &    &   & (osx4) \\
 \hline

bias  $\langle \REin - {\REin}_{0}\rangle$  & $-0\farcs040$  & $0\farcs078$ &  $-0\farcs014$   & $0\farcs048$ \\
scatter $ \sigma_{ \REin - {\REin}_{0} }$  &  $0\farcs078$  & $0\farcs257$ &   $0\farcs043$  & $0\farcs220$ \\
bias  $\langle \log \frac{\mu}{\mu_0}\rangle$  & $-0.105$   & $-0.080$ &  $-0.018$  & $-0.041$ \\
scatter $ \sigma_{ \log(\mu/\mu_{0}) }$  & $0.125$   & $0.344$ & $0.084$  & $0.240$ \\

bias  $\langle m_{\rm img} - m_{\rm img,0}\rangle$  & $-0.070$   & $2.074$ &  $0.020$  & $0.304$ \\
scatter $ \sigma_{ m_{\rm img} - m_{\rm img,0} }$  &  $0.111$  & $1.048$ &  $0.119$  & $1.402$ \\ 
cat. $|\Delta\REin| > 10\,\delta\REin$  & 38\% & 34\% & 14\% & 41\% \\  \hline
\% failures &  9\%  &14\% &  16\%  & 31\% \\
median exec. time &  10 s  & 8 s &  18 s  & 5 m 42 s \\
\hline
\end{tabular}
\tablefoot{Distinguishing the 4 different cases considered in \S\ref{sec:resmock:ideal} and \S\ref{sec:resmock:galfit} (ground-based vs space-based simulated observations, and ideal vs \galfit deflector subtraction), we report in this table the mean bias and the scatter observed for the 3 main parameters of the lens model ($\REin$, $\mu$, $m_{\rm img}$), along with the rate of catastrophic errors defined as the systems for which $|\Delta\REin|/ \delta\REin> 10$, knowing that the errors are underestimated by the fast modeling. Concerning the time aspect, we give the median execution time and the rate of failures corresponding to the cases where the MCMC did not converge within a threshold time relative to the median time (1 minute for \cfht simulations whatever the subtraction method, 2 minutes for \hst simulations with an ideal subtraction, and 45 minutes for \hst simulations with a \galfit subtraction due to the necessary $\times4$ sampling while simulating and optimizing.) }
\end{table}
 
The recovery of the total magnification shows two clearly separated regimes : on one hand, an overestimation regime with a high scatter, on the other hand, a sharp underestimation regime with a small scatter, roughly corresponding to the respective overestimation and underestimation of $\REin$. For \hst data, those two modes are significantly alleviated, and for both sets of data, the median value of $\mu$ is not so biased either. Concerning the total flux of the lensed light, the way it is recovered somehow echoes the faulty recovery of the deflector flux with \galfit : contrary to $\REin$ and $\mu$, $m_{\rm img}$ is thus significantly biased, even for \hst data. 

We defined the catastrophic errors as the optimizations satisfying $|\Delta \REin| > 10\,\delta\REin$, the errors being quite underestimated by the fast modeling (see \S\ref{sssec:optim}). With an ideal deflector subtraction, the rate of catastrophic errors is 14\% for \hst simulations and 38\% for \cfht simulations. In the case of a \galfit subtraction, this rate remains very similar to its previous value for \cfht simulations but reaches 41\% for \hst simulations. We tested that the oversampling of the model-predicted light distribution is needed for \hst simulations as it allows the number to catastrophic errors to decrease from 54\% to 41\%. More precisely, we observe that low signal-to-noise, low magnification, low values of ${\REin}_0$ lead to slightly smaller fractions of failures (because of the increased formal error $\delta\REin$), but the dependency of the fraction on either ${\REin}_0$, $\mu_0$ or $m_{\rm img,0}$ is mild, especially for HST conditions. The dramatic reshuffling of the systems in the recovered parameter space with \galfit subtraction washes those trends out.

 Coming back to less magnified lensing configurations at large radii, besides an increased scatter (already mentioned for the case of an ideal deflector subtraction) and a few outliers, conclusions stay unchanged since multiple image systems still constitute a rather favorable situation for \galfit when the Einstein radius is large, the main arc remaining sufficiently far away from the deflector.


For ground-based simulations, the median time for the optimization to complete for the subtraction with \galfit is similar to the minimization with ideal subtraction, that is $\sim 8\,$s. However, 14\% of the minimizations take more than 1 minute to converge as compared to $\sim9\%$ in the previous case, and among them more outliers have emerged. This increased failure rate is due to the complex shape of the posterior distribution function often induced by the faulty foreground subtraction.\\
For simulations of space-based data, because of the finer sampling of the light distribution predicted by the model, the median execution time rises to $\sim6$ minutes. The failure rate reaches 31\% for a threshold value of 45 minutes.

\section{Calibrating \slfit as an automated lens finder with mock data}\label{sec:resmock:robot}

Having shown that \slfit  gives an accurate estimation of the most important parameters characterizing a galaxy-scale
strong lensing event, we now keep on with the idea of using the \slfit output of a given lens model
as a way to decide whether it is a gravitational lens or not. All our previous lens modeling assumptions, and in particular the use of \galfit to get rid of the deflector's light, are kept unchanged. As already said, we expect that the faulty \galfit subtraction will be a source of concern in the lens finding objective. So we account for this limitation here. From the perspective of using \slfit as a lens finding algorithm on current and upcoming ground-based datasets, we do not need to discuss further the \hst simulations and focus on the CFHT case.

In the previous section, we wanted to assess the reliability of \slfit at recovering the main lensing parameters that are the Einstein radius, the total magnification $\mu$ and the total magnified flux $m_{\rm ing}$ of interesting lenses over a broad range of values regardless of their actual frequency in realistic lensing configurations. To do so, we restricted ourselves to a very small impact parameter for the source $\beta \le 0\farcs2$. However, this simulation setup does not reproduce the relative abundance of low and high magnification configurations, nor it uniformly samples the source plane domain over which \slfit might yield a positive signal. We therefore had to simulate another 3000 systems in the range $0\farcs2 < \beta \le 2\farcs5$ on top of the already simulated 1260 systems with $\beta\le 0\farcs2$. Hence, the source plane is sampled with two levels of sparsity : a system with a source inside the inner circle will be given a statistical weight $w_{\rm in}= 0.2^2 / 2.5^2 \times 3000/1260 \simeq 0.015$ while sources inside the outer annulus are given a weight $w_{\rm out}=1- 0.2^2 / 2.5^2 \simeq 0.99$.  Besides those large impact parameter, low magnification systems, we also need to account for chance alignment systems that do not give rise to high magnification but could be flagged by \slfit output as potential lenses.  Systems in which the assumed source actually lies in front of the assumed deflector are therefore included in the statistical analysis of \slfit output.

To set up the selection, one first needs to classify the simulated lens candidates according to their true characteristics. Following G14, we consider as potentially interesting systems all the previously simulated lines of sight that produce a total magnification $\mu_0>4$ (good lenses), and consider as quite interesting systems having $2\le \mu \le 4$ (average lenses).
Moreover, as in the previous section, we will not aim at uncovering lenses for which the magnified arc or ring has a total image plane magnitude $m_{\rm img,0}> 24.5$ in the CFHT $g$ band.

We want our decision process to be simple and we therefore design a binary decision (lens or not-a-lens) based on  \slfit ``best-fit'' parameter estimates defined as the median values of the marginal posterior distributions (see \S~\ref{sec:model:infer}). Conservative requirements on the quality of the optimization result are also introduced by applying a simple threshold on the reduced $\chi^2/\nu$ statistics. Since the MCMC errors are relatively underestimated by the fast modeling, we do not take them into account.

\begin{figure}[h]
\centering
          
	   \includegraphics[width=8.5cm]{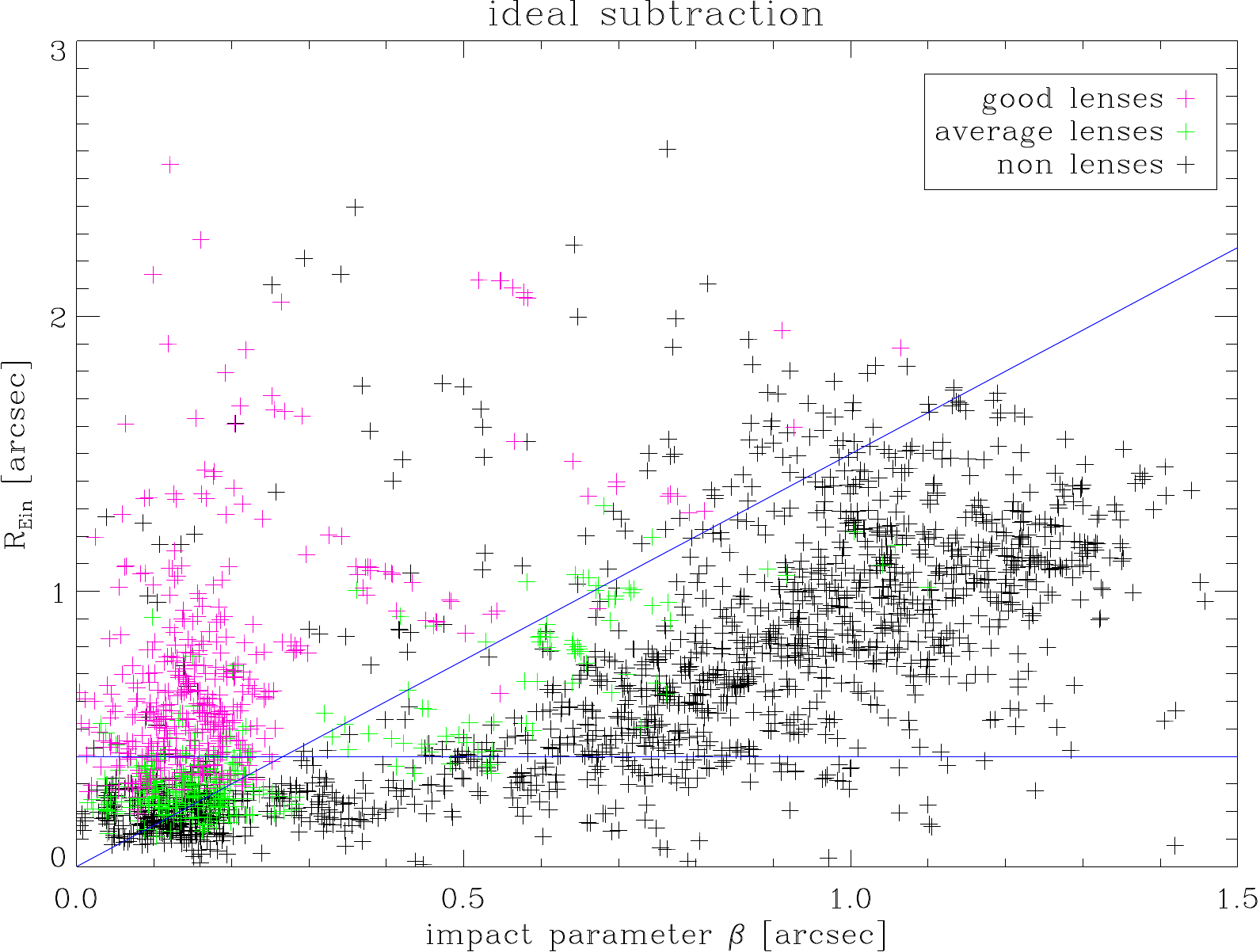}  \\
	   \vspace{0.5cm}
	   \includegraphics[width=8.5cm]{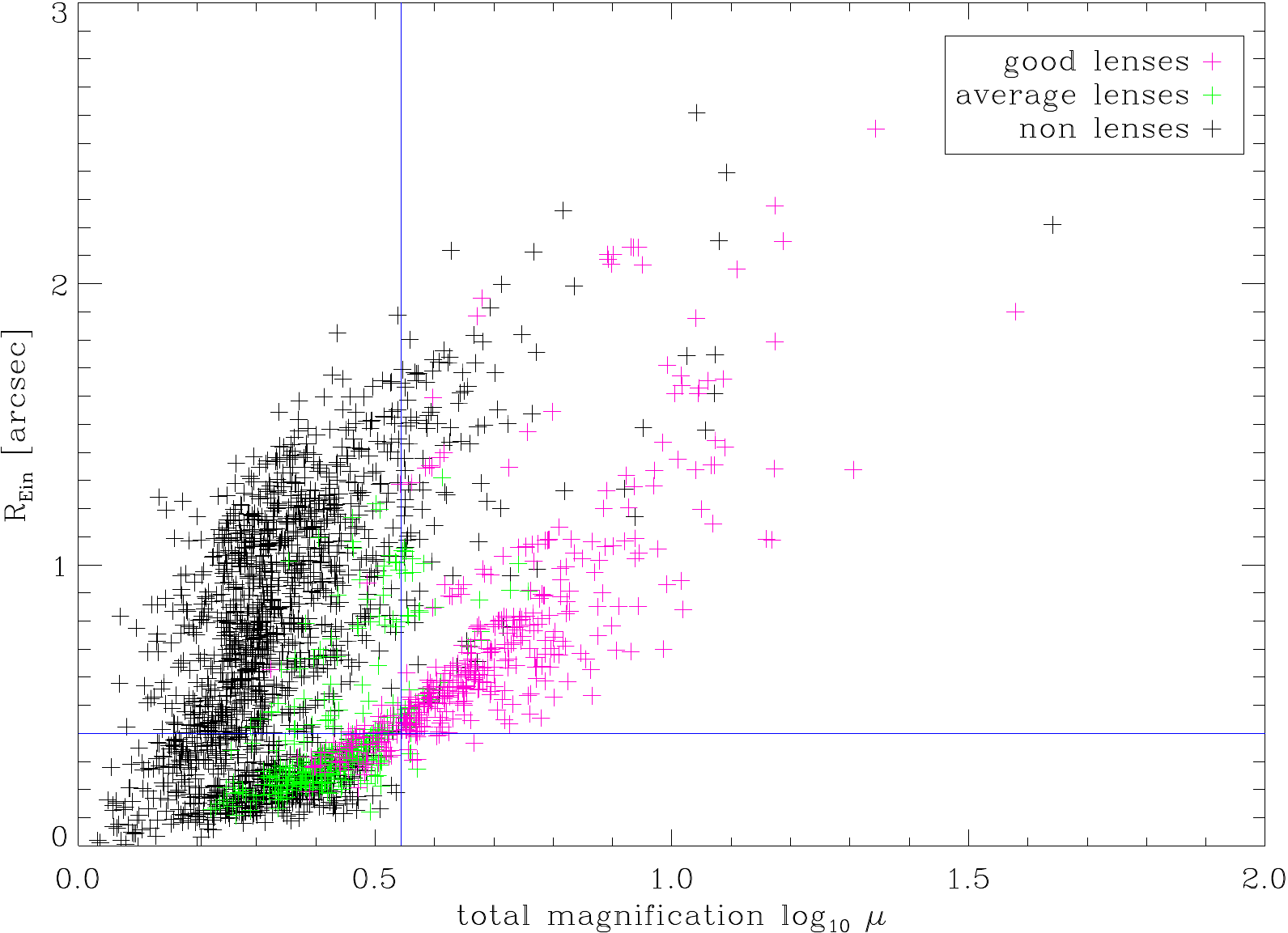}  \\
	   \vspace{0.5cm}
	   \includegraphics[width=8.5cm]{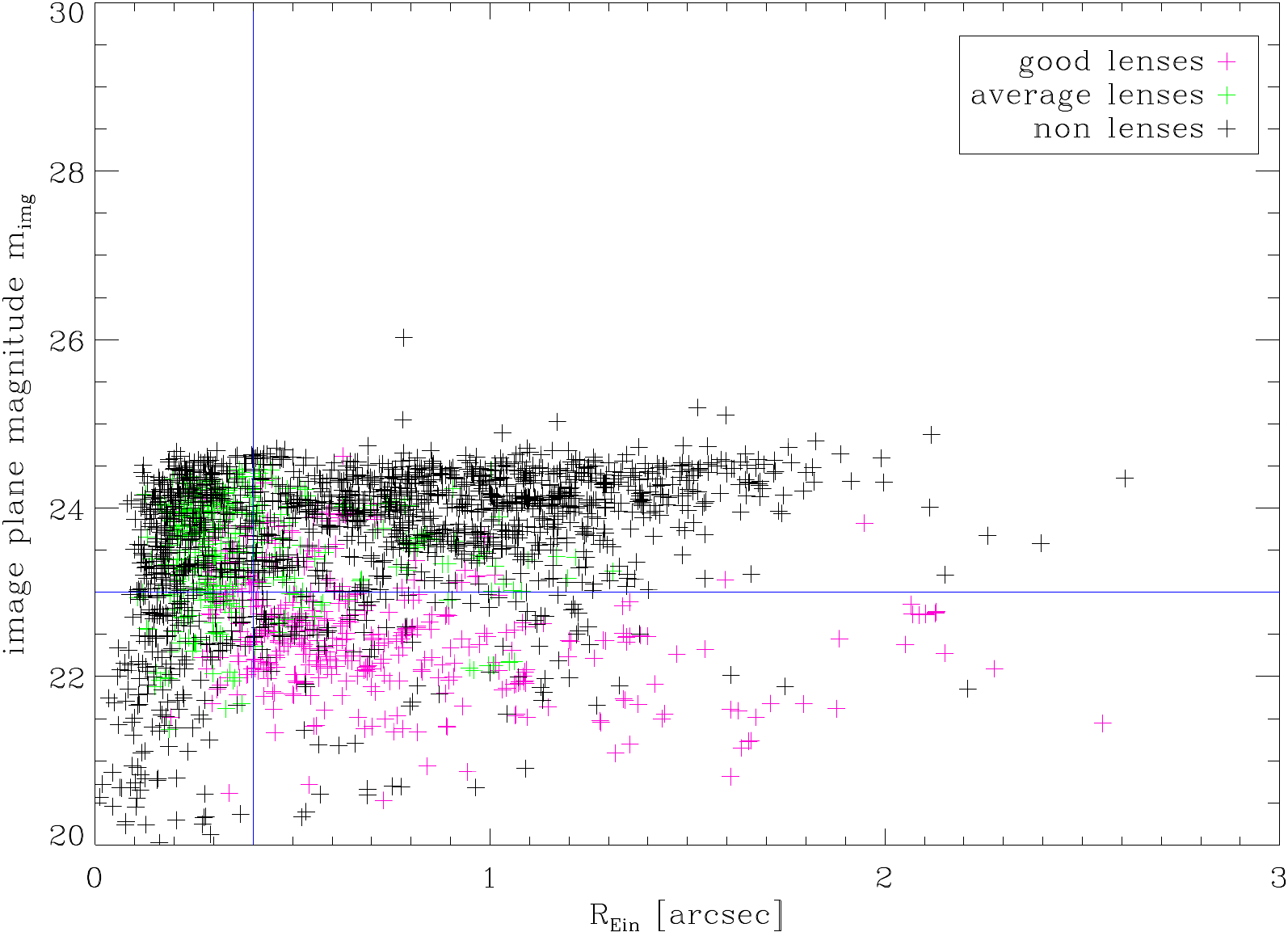}
	   	   \caption{Results of the \slfit modeling in the \{$\REin$, $\beta$\} \textit{(top)}, \{$\REin$, $\mu$\} \textit{(middle)} and \{$\REin$, $m_{\rm img}$\} \textit{(bottom)} subspaces for an ideal subtraction of the deflector. Three classes of lenses are considered : 'good lenses' defined by $\mu_{0}>4$   \textit{(pink)}, 'average lenses' such that $2<\mu_{0}<4$ \textit{(green)}, and non lenses \textit{(black)}. Lines represent the empirical cuts applied to isolate lenses. They are defined in Eq.~\eqref{eq:cutsID}.}\label{fig:mockselec}
\end{figure}

Fig.~\ref{fig:mockselec} shows the distribution of the simulated and fitted lens candidates in the \{$\mu$, $\REin$, $m_{\rm img}$, $\beta$\} parameter space when the light of the foreground deflector has been ideally subtracted off, with a color distinction for 'good', 'average' and 'non' lenses. Very simple cuts of the kind:
\begin{equation}\label{eq:cutsID}
\begin{aligned}
 & \{\,\REin > 0\farcs4,\;\; \REin > 1.5\,\beta,\;\; \mu> 3.5,\;\; m_{\rm img} < 23,\;\; \chi^{2}/\nu < 2\,\}
\end{aligned}
\end{equation}
are able to isolate the most interesting lenses. We can therefore define the completeness as the number of $\mu_0>4$ systems satisfying \eqref{eq:cutsID} divided by the total number of $\mu_0>4$ 'good' systems. Likewise, we define the purity as the number of $\mu_0>4$ systems satisfying the criterion \eqref{eq:cutsID} divided by the total number of systems satisfying the same criterion. We used this simulated sample of lens candidates to maximize the purity as a first step while preserving completeness: Table~\ref{tab:resrobotmock} gathers those results.

\begin{figure}[h]
\centering
       	\includegraphics[width=8.5cm]{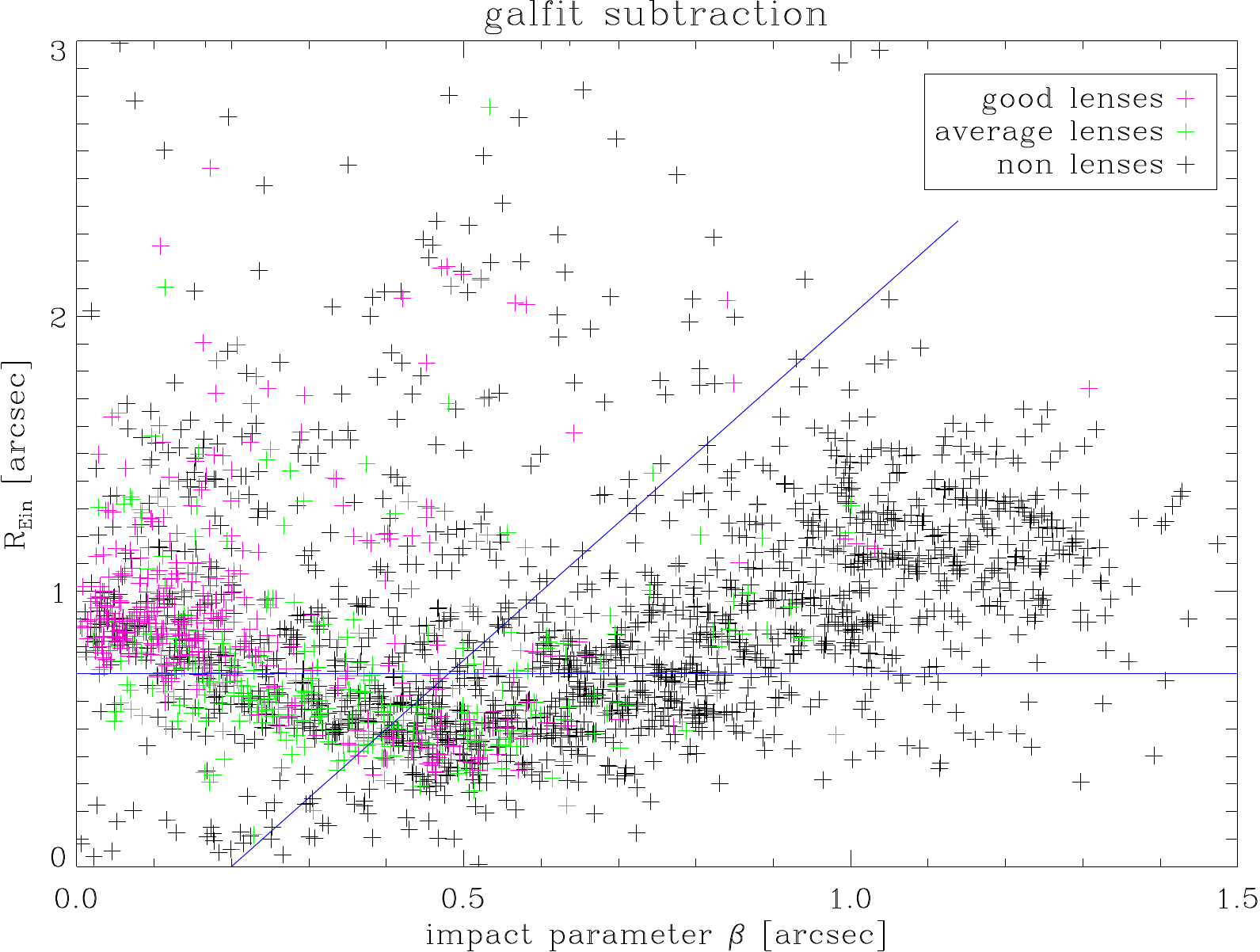} \\
	\vspace{0.5cm}
	\includegraphics[width=8.5cm]{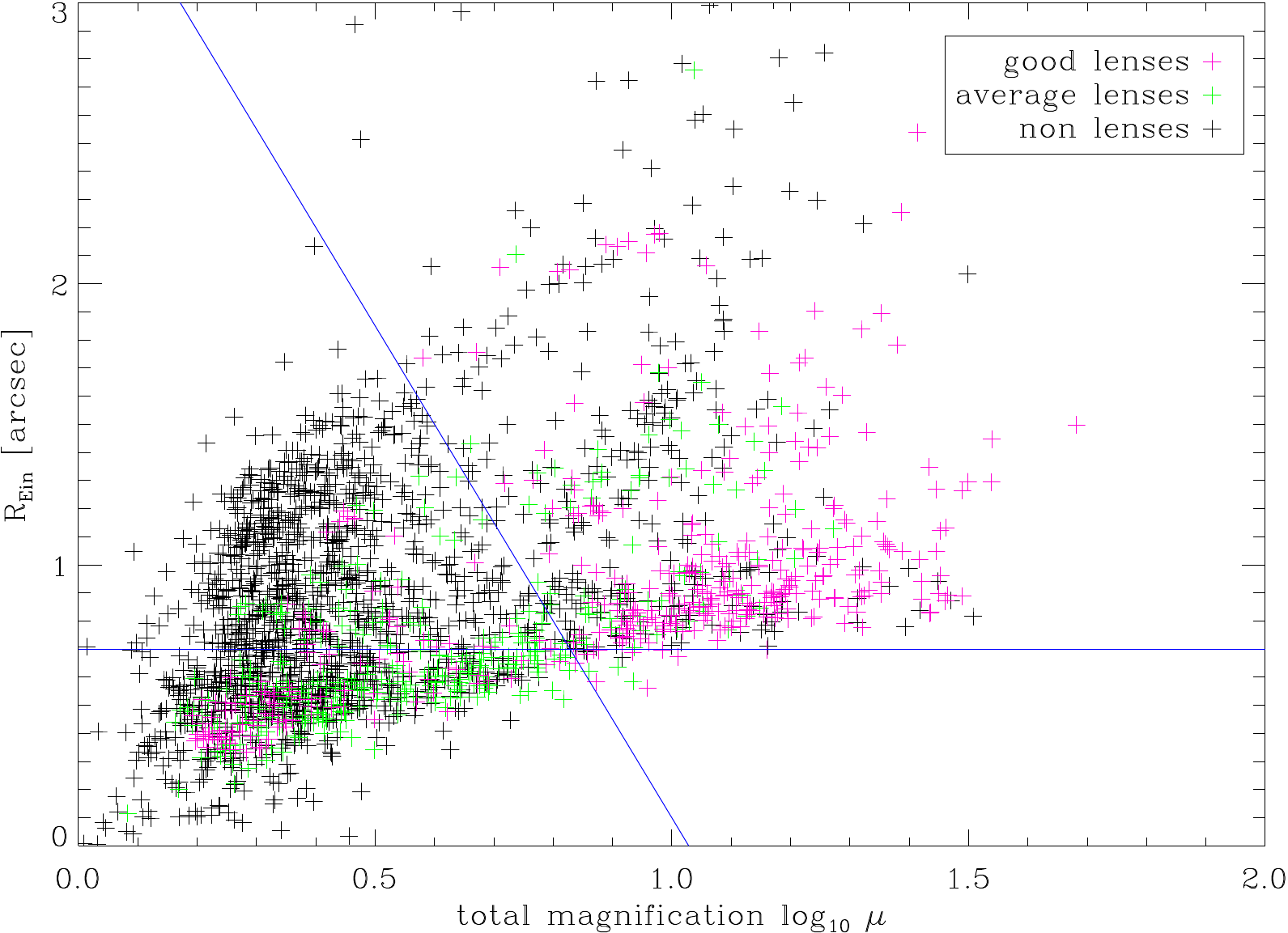} \\
	\vspace{0.5cm}
	\includegraphics[width=8.5cm]{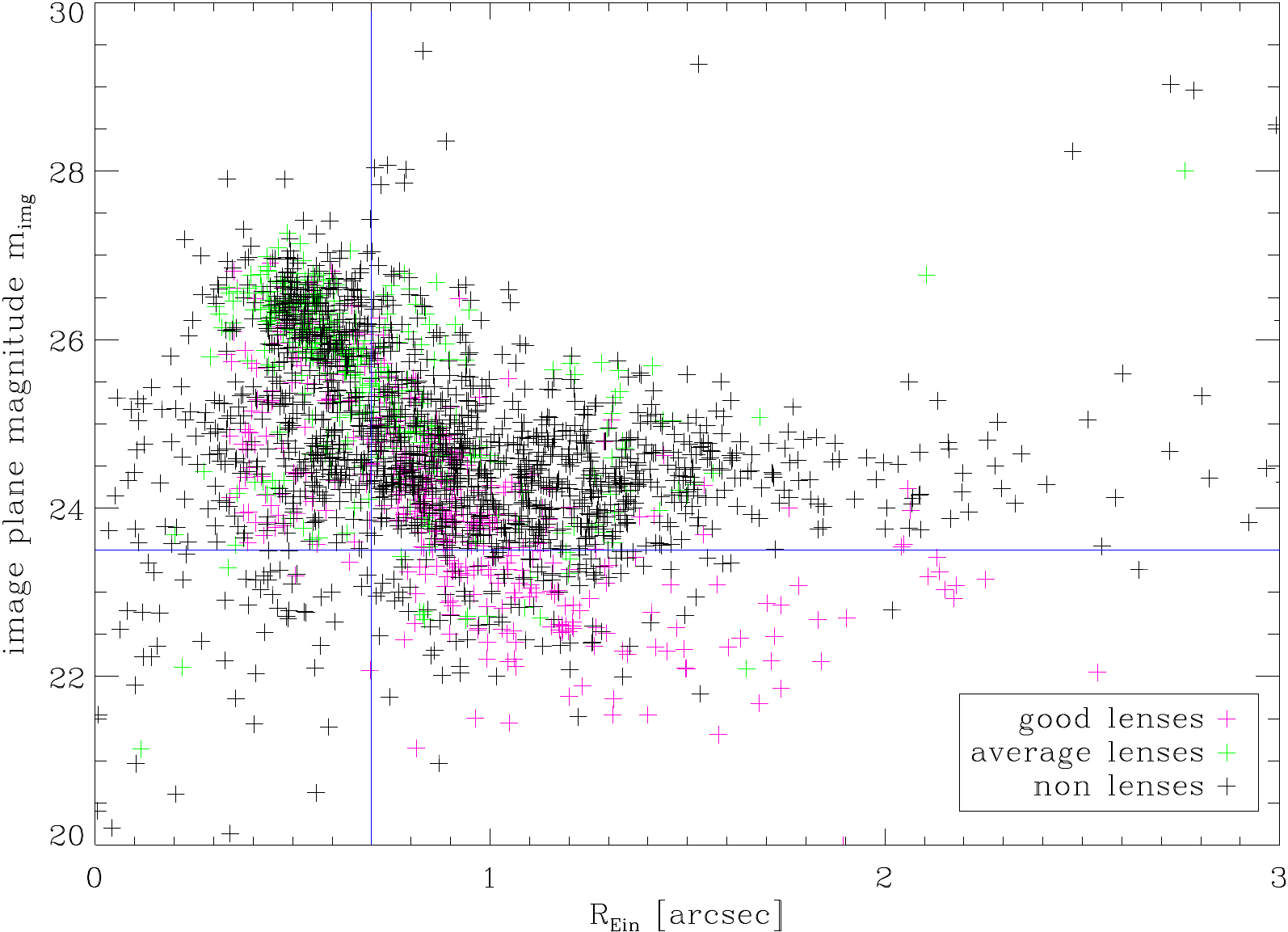}
	   \caption{Same as Fig.~\ref{fig:mockselec} but for a \galfit subtraction of the deflector. Lines represent the cuts defined in Eq.~\eqref{eq:galcutsID}.}\label{fig:mockselecGF}                    
\end{figure}

In the more realistic case of a \galfit subtraction, it is much more difficult to isolate the 'good' lenses from the 'non' lenses as can be seen in Fig.~\ref{fig:mockselecGF}. The best tradeoff between completeness and purity is now more difficult to achieve. We came up with the following cuts:
\begin{equation}\label{eq:galcutsID}
\begin{aligned}
&\;\;\;\;\; \;\;\;\;\; \;\; \;\;\{\,\REin > 0\farcs7,\;\;\;\;\; \REin > 2.5\,\beta-0\farcs5,\\
&\REin > -3.5\,\log{(\mu)}+3\farcs6, \;\;\;\;\; m_{\rm img} < 23.5,\;\;\;\;\;\chi^{2}/\nu < 10\,\}
\end{aligned}
\end{equation}
for which the results are also listed in Table~\ref{tab:resrobotmock}.

We get very good results selecting 'good' lenses in the case of an ideal subtraction, ending up with 82\% of completeness and 94\% of purity. Unfortunately, due to the imperfect \galfit subtraction spotted in \S\ref{sec:resmock:galfit} that tends to mix up in the parameter space the different areas corresponding to the different classes of lenses, the completeness and purity are substantially reduced to 31\% and 69\% respectively. Considering individually 2-D slices through the involved parameter space, there is an important degree of overlap between the selections resulting from the sets of cuts applied to \{$\REin$, $\beta$\} on the one hand, and to \{$\REin$, $\mu$\} on the other hand, but we decided to consider both selections here in our attempt to reach a high level of purity. Furthermore, in the case of an ideal subtraction, those two first sets of cuts are at the same time more complete and more selective than the cuts made through \{$\REin$, $m_{\rm img}$\} only. 
When switching to a \galfit subtraction, we find that the selections derived from \{$\REin$, $\beta$\} and \{$\REin$, $\mu$\} remain more discriminating than the one derived from \{$\REin$, $m_{\rm img}$\}, but that the latter is more complete than the second one; furthermore, inside the sample, the selection involving $m_{\rm img}$ is also more complete for systems with  $\beta > 0\farcs2$ since, as noted before, \galfit systematically assigns a substantial fraction of the lensed flux to the foreground unlensed light distribution when $\beta$ is small. 
 Regarding 'average lenses', they are already quite mixed with non lenses for an ideally-subtracted deflector, so we did not quantify the completeness and purity relative to that class of lenses. Finally, knowing that we have favored the purity rate, the results remain relatively satisfactory even for the \galfit subtraction; yet, the loss is substantial as compared to the ideal case, showing that the method chosen for the deflector subtraction plays a crucial role in the lens modeling and in the subsequent selection process.
\begin{table}[ht]
\centering
\caption{\label{tab:resrobotmock} Selection results on the simulated lens candidates}
\begin{tabular}{lcc}\hline\hline
 Subtraction type &  ideal  & \galfit    \\ \hline
Completeness & 82\% & 31\% \\ 
Purity & 94\% & 69\% \\   \hline
 \end{tabular}
\tablefoot{Completeness (rate of targeted lenses that have been selected from all the targeted lenses in the initial sample) and purity (rate of targeted lenses in the selected sample) for the 'good lenses' ($\mu_{0}>4$) resulting from the selection of Eq.~\eqref{eq:cutsID} for an ideal subtraction of the deflector, and of Eq.~\eqref{eq:galcutsID} for a \galfit subtraction.}
\end{table}



\section{Application to real CFHTLS data}\label{sec:resreal}
We now apply the automated model fitting  to the large CFHTLS dataset, whose statistical properties are  similar to simulated images.  Although we leave a fully automated analysis of the entire survey for a future work, here we make the most of the existing sample of strong lenses built by G14 in order to test the reliability of the above conclusions.
We can hence use a large sample of confirmed lenses and good quality lens candidates to assess the relative completeness between our new method and the previous \RF algorithm. In addition, real data will bring into the problem several important new ingredients that are missing in the simulations, like a complex morphology of the foreground deflector and background source but also the environment of potential lenses that is more complex due to the presence of neighboring objects that could interfere in the light fitting process either at the deflector subtraction stage or at the subsequent stage of fitting the residual potentially lensed light.

\subsection{The SL2S \RF sample of strong lenses found in the CFHT-LS}\label{ssec:cfhtls}

We take advantage of the existing SL2S sample \citep{Ruf++11,Gav++12,PaperIII,PaperIV,PaperV,Gav++14} of candidate and confirmed galaxy-scale strong lensing systems found in the CFHTLS with the \RF algorithm. The goal is to investigate the response of our automated modeling method to a large number of interesting lens candidates.

The Canada-France-Hawaii Telescope Legacy
Survey\footnote{\url{http://www.cfht.hawaii.edu/Science/CFHLS}}
(CFHTLS) is a major photometric survey of more than 450 nights over 5
years (started on June 1st, 2003) using the MegaCam wide field imager
which covers $\sim$1 square degree on the sky, with a pixel size of
0\farcs186. The CFHTLS has two components aimed at extragalactic
studies: a Deep component consisting of 4 pencil-beam fields of
1\,deg$^2$ and a Wide component consisting of 4 mosaics covering
150\,deg$^2$ in total. Both surveys are imaged through 5 broadband filters. 
In this paper we use the sixth data release (T0006) described in detail by
\citet{Gor++09}\footnote{see also,
\url{http://terapix.iap.fr/rubrique.php?id\_rubrique=259},
\url{http://terapix.iap.fr/cplt/T0006-doc.pdf}}
The Wide survey reaches a typical depth of $ u^*\simeq 25.35$, $
g\simeq 25.47$, $r\simeq24.83$, $i\simeq 24.48$ and $ z\simeq 23.60$
(AB mag of 80\% completeness limit for point sources) with typical
FWHM point spread functions of $0\farcs85$, $0\farcs79$, $0\farcs71$,
$0\farcs64$ and $0\farcs68$, respectively. Because of the greater
solid angle, the Wide survey is the most useful component of the CFHTLS for finding and studying strong lenses.
Regions around the halo of bright saturated stars, near CCD defects or
near the edge of the fields have lower quality photometry and are
discarded from the analysis. Overall, $\sim 21\%$ of the
CFHTLS Wide survey area is rejected, reducing the total usable area to
135.2 deg$^2$.

The \RF~lens finding method is presented in G14. This algorithm
detects compact rings around  isolated galaxies, and works by looking for blue
features in excess of an early-type galaxies (ETGs) smooth light
distribution that are consistent with the presence of lensed arcs.
After selecting a sample of bright ($i_{\rm AB}
\le 22$) red galaxies,  a scaled, PSF-matched
version of the $i$-band image was
subtracted from the $g$-band image. The rescaling in this operation is
performed such that the ETG light is efficiently removed, leaving only
objects with a spectral energy distribution different from that
of the target galaxy. These typically blue residuals are then
characterized with an object detector, and analyzed for their
position, ellipticity, and orientation, and those with properties
consistent with lensed arcs are kept as lens candidates. In
practice, \RF requires the blue excess to be elongated (axis ratio
$b/a<1/2$) and tangentially aligned ($\pm 25^\circ$) with respect to
the center of the foreground potential deflector. The objects
are searched for within an annulus of inner and outer radius
$0\farcs5$ and $2\farcs7$, respectively. The lower bound is chosen to
discard fake residuals coming from the unresolved inner structure of
the deflector, inaccurate PSFs, etc. The outer bound is chosen to
limit the detection of the many singly imaged objects (see G14). 
In practice, \RF finds about 2-3 lens candidates per square degree. Extensive follow-up
shows that the sample is 50\%-60\% pure (G14).

This sample with a large proportion of actual strong lensing system is an ideal training set for our automated lens finding method. In addition, the abundant follow-up observations achieved by the SL2S team also provides accurate lens models based on high-resolution imaging for a fraction of the sample \citep{PaperIII}. This can be used to test the reliability of a automated lens modeling procedure when run on real CFHTLS data. We shall thus apply the same modeling technique to a sample of 517 SL2S lens candidates (including both confirmed lenses, a few confirmed non-lenses, and some still-to-be-confirmed systems).
In addition, in order to have a better handling on the purity of the sample another 305 ETGs are also 
randomly picked in the CFHTLS to enhance the population of modeled non-lenses and better control the response of \slfit in the presence of actual non-lenses.

\subsection{Fitting SL2S lens candidates with \slfit}
Before addressing in more detail the ability of our automated lens modeling technique to test the lens / not-a-lens nature of a given galaxy, we need to assert that \slfit and the specific modeling assumptions we have made for this project (cf \S\ref{sec:model:assump}) are still suitable for real data.  We also take advantage of the \citet{PaperIII} models of some of the SL2S lenses that were performed on HST and CFHT images to check the fidelity of the fast modeling strategy.

 \subsubsection{New modeling issues affecting real data} \label{subsubsec : new issues}
All the diversity and complexity of actual observed lenses cannot be easily incorporated into simulations or the modeling assumptions. It affects the subtraction of the deflector by \galfit, and the subsequent fit of the lens potential and of the source by \slfit. So one can expect final models to differ from the truth if, e.g., the background source is not a perfect exponential profile with an elliptical shape or if the foreground deflector is not a perfect de Vaucouleurs  profile but rather exhibits some additional light at its periphery, or if the actual lens potential is far from a simple SIE model. Galaxies are not perfectly isolated and neighbors, either in projection or physically correlated, may perturb the fitting of the light distribution of a given galaxy. If some sources happen to be close to the fitting region, one will have to mitigate its effect on the modeling of the lensing galaxy and its lensed features as they could severely drive the shape of the posterior function and produce undesired local minima since the lensed features are generally quite dim.
\begin{figure}[htb]
\begin{multicols}{2}
  \includegraphics[width=4.45cm]{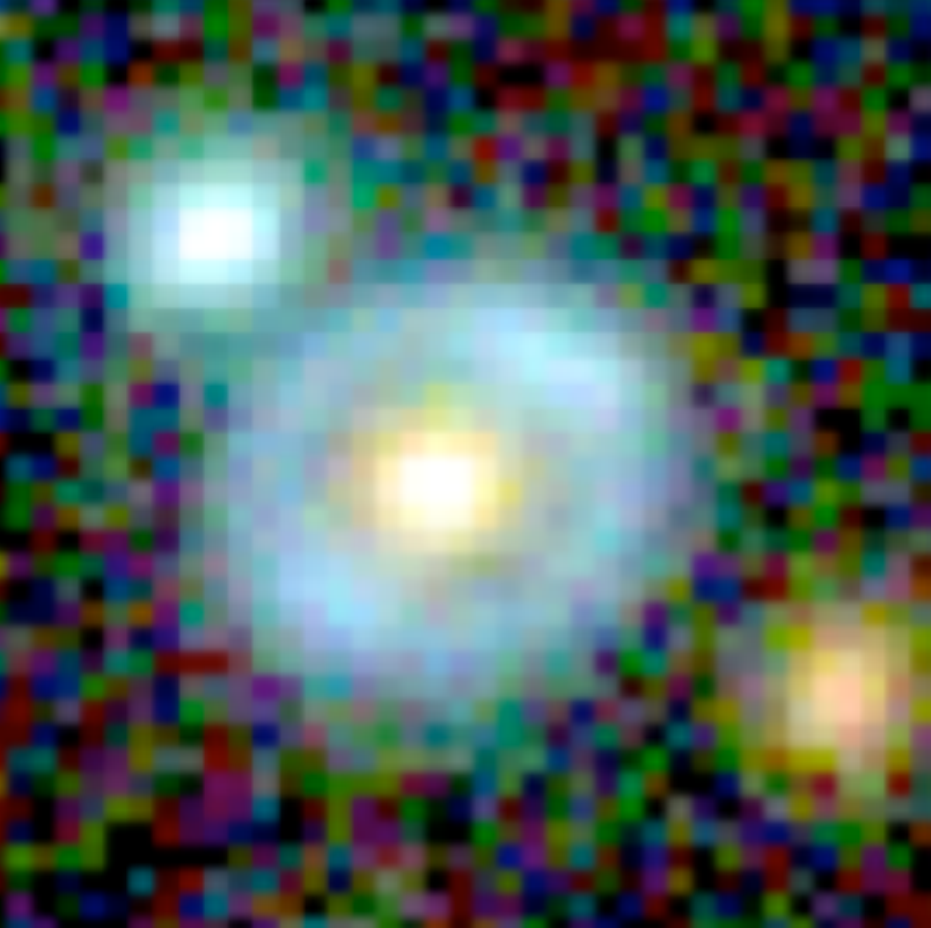} 
  \columnbreak  \\
  \includegraphics[width=4.45cm]{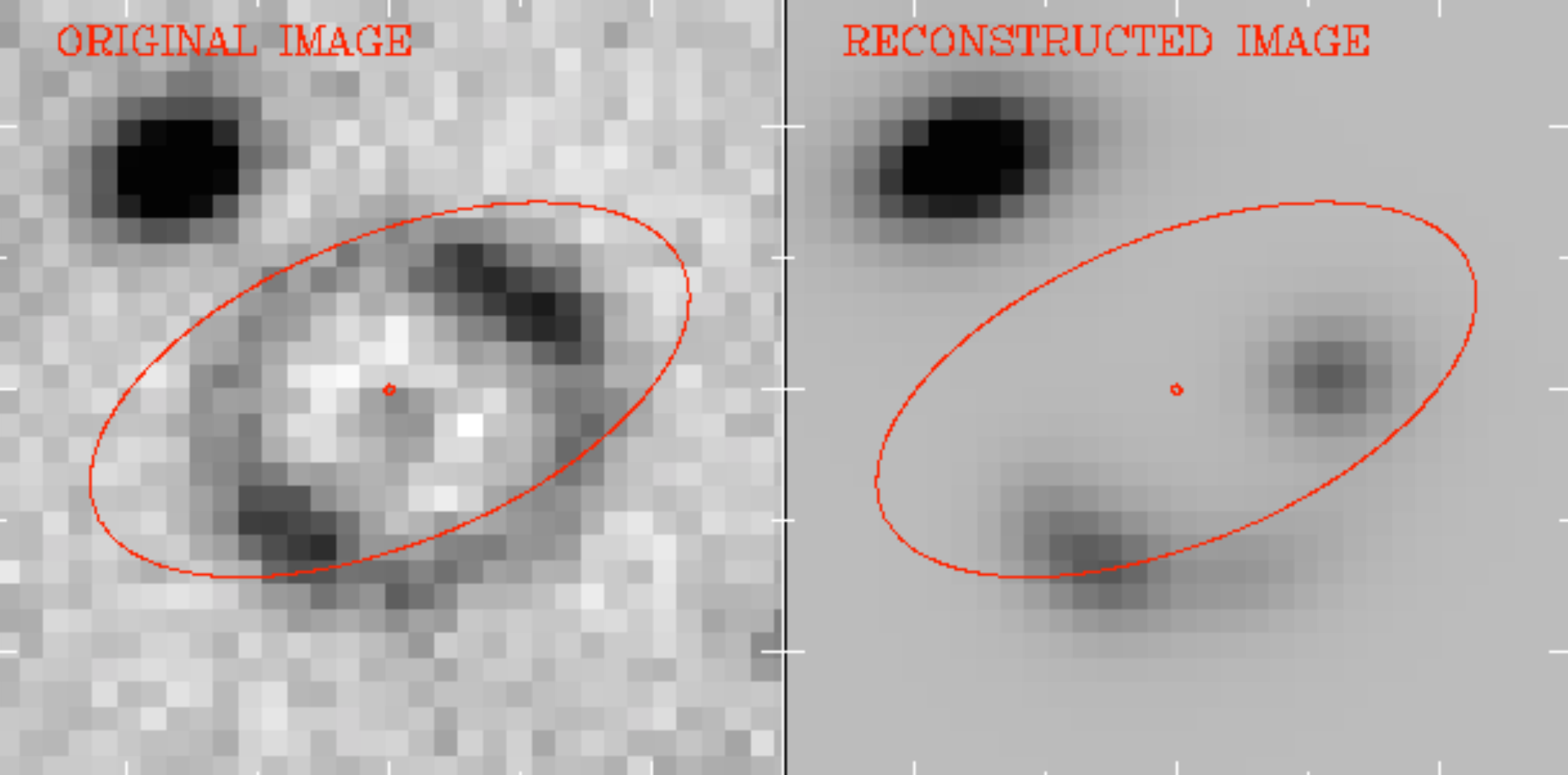}
  \includegraphics[width=4.45cm]{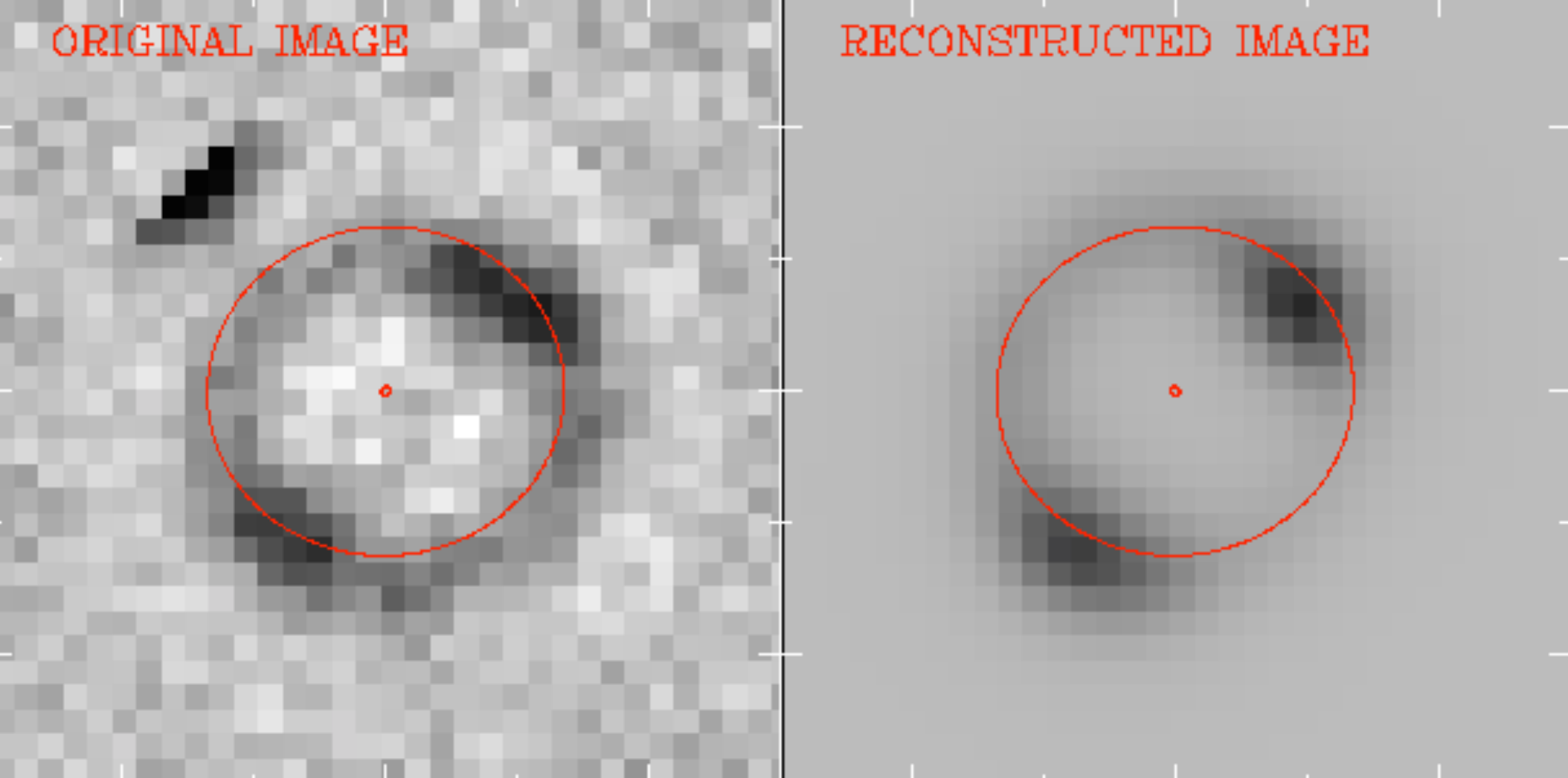}
                                   \end{multicols}
\vspace{-0.75cm}
\caption {Blue ring from the CFHTLS with a neighboring galaxy of a similar color in the observed projected plane \textit{(left panel)}. Without mask, \slfit identifies this piece of environment as a bright lensed feature and looks for a wrong lensing configuration \textit{(top right panel)}. Here, our standard masking eliminates most flux from the blue galaxy and thus helps the modeling to focus on the actual lensed flux \textit{(bottom right panel)}.  As in Fig.~\ref{fig:compmodel}, we show in each case the original image where the deflector has been subtracted in the $g$ band, and the gravitational arcs or ring reconstructed by the model with the critical lines \textit{(in red)}.}\label{fig:bluering}
\end{figure}

In order to mitigate these  nuisances, we masked out of the input image all pixels at distance less than  $0\farcs3$ and beyond $2\farcs5$. We hence assume that the multiple images (arcs, rings) nestle inside this thick annulus. This masking is performed after the deflector subtraction by \galfit and only concerns the subsequent \slfit fitting. Like for simulated data, the fitting window is $7\farcs7$ on a side.
We chose to replace pixel values with pure noise realizations of zero mean and variance given by the input weight map. Hence lensed features appearing beyond $2\farcs5$ will be chopped off and not properly modeled. As an alternative, we explored the possibility to just neglect those masked pixels in the $\chi^2$ calculation, which would not prevent the model from producing multiple images beneath the masked area.
Moreover, by carrying out the fitting on a smaller number of pixels, the convergence of the chain is expected to happen faster. Yet there is a risk here of producing models far from the physical reality of some systems by conceiving arcs beyond $2\farcs5$ that do not actually exist. Our experience suggested that the former approach was preferable if the goal is to avoid producing too many false large $\REin$ systems. Fig.~\ref{fig:bluering} shows the example of an observed ring from the CFHTLS for which the application of this simple masking is successful at recovering a lens configuration much closer to the truth . 

\subsubsection{Comparison with published models of SL2S lenses}
\citet{PaperIII} produced detailed lens models of several confirmed SL2S lenses that had been observed with Keck or VLT spectrographs to measure redshifts and stellar velocity dispersions \citep[see also][]{PaperIV,PaperV}. Some systems were analyzed with \hst imaging and others were relying on the same \cfht data as here. The main difference resides in the level of refinement devoted to foregound light subtraction, to the accounting of neighboring objects and to a much richer model for the background source which is pixelized, all things that we cannot currently afford within a fast modeling strategy. We can therefore assume that \citeauthor{PaperIII} models are somehow "the truth'' for our present goal. The subscript "Son++" refers to the parameter values from these models.

\begin{figure}[ht]
  \centering
    \includegraphics[width=8.5cm]{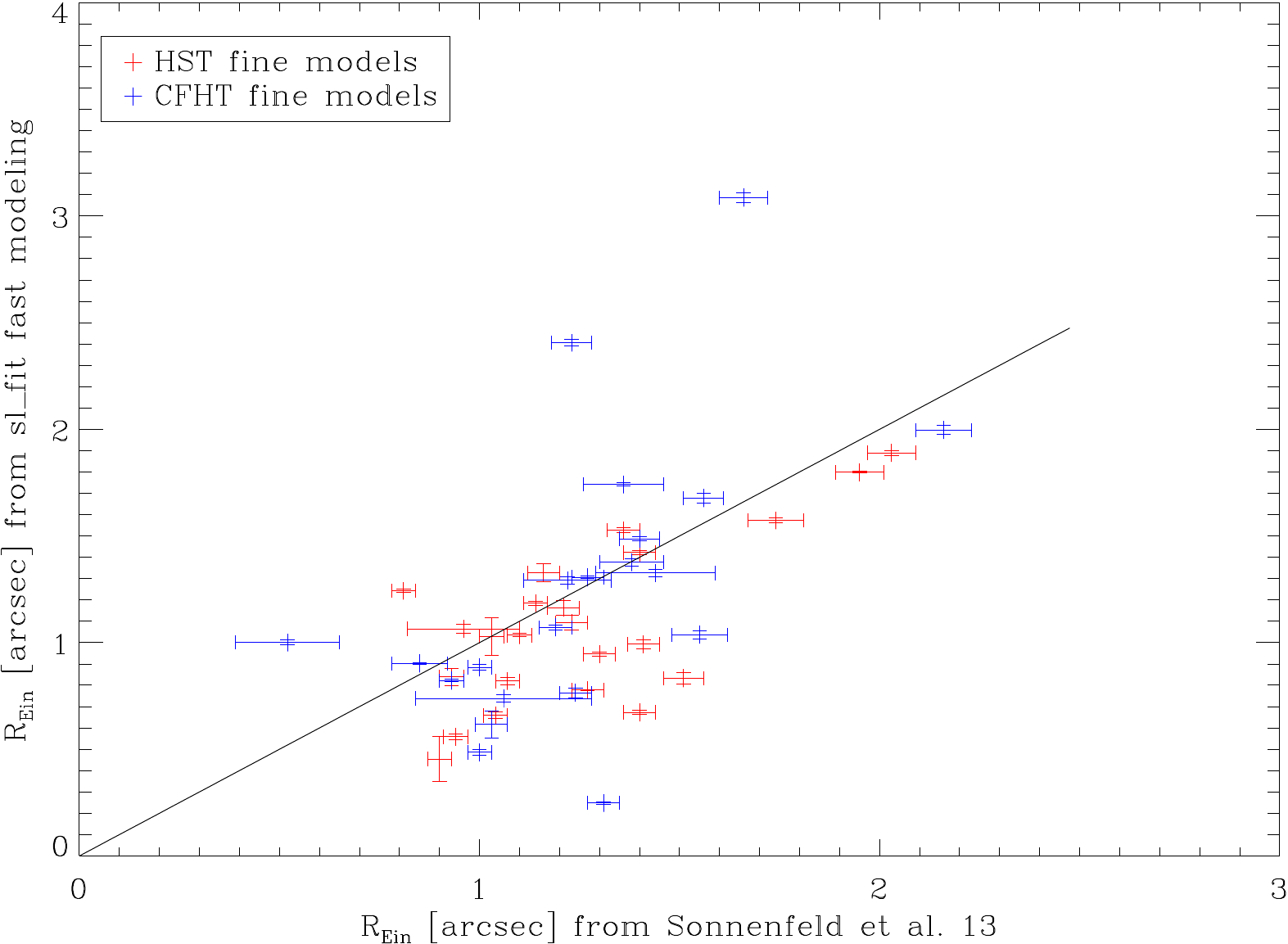}
  \caption{Comparison between the Einstein radius $\REin$ measured by \citet{PaperIII} and our fast \slfit inferences for 44 confirmed SL2S lenses. \citeauthor{PaperIII} models relying on \hst (red) and \cfht (blue) images are distinguished.}\label{fig:comp47}
\end{figure}

Fig.~\ref{fig:comp47} compares both estimates of $\REin$ for a sample of 44 systems out of the 517 SL2S candidates that we modeled. Here, we have excluded a few cases that appeared to be especially complex, the observations suggesting peculiarities like a double deflector, a composite source hosting an AGN, a very bright piece of environment etc. 
Moreover, the mask we decided to automatically apply to every candidate (see \S~\ref{subsubsec : new issues}) affects lens systems with  ${\REin}_{\rm Son++} \gtrsim 2\farcs 5$ by hiding substantial flux belonging to one or more lensed arcs, so we focused on   systems such that ${\REin}_{\rm Son++} < 2\farcs 5$ for this comparison.

The scatter in the ${\REin}  \leftrightarrow  {\REin}_{\rm Son++}$ relation for \hst models, given by $\sigma_{ {\REin}_{\rm Son++} - {\REin} } \simeq 0\farcs28$, is  close to the $\sim0\farcs26$ value found for the simulations : this somehow proves the robustness of our $\REin$ estimate despite the new obstacles brought by real data. For \cfht data, the scatter is a bit larger $\sim0\farcs55$. This   value is mostly driven by 3 outliers for which the complete lack of visible counter image led the optimization to an unlikely lensing configuration : without those 3, we get the same satisfactory $\sim0\farcs28$ value as for \hst models.

\subsection{Calibrating \slfit as an automated lens finder with real CFHTLS data: selection strategy}
In \S\ref{sec:resmock:robot}, we have used a sample of 4260 simulated lens candidates of different classes, including non lenses, so as to maximize the purity rate of our selection while maintaining a reasonable completeness rate. For a \galfit subtraction, the only previously addressed case that can be reproduced on real data, we ended up with a purity rate of 69\% and a rate of completeness of 31\%. Using the preselected  sample of 517 SL2S lens candidates, we now want to recalibrate completeness and purity depending on the fraction of 'real good lenses' that can be actually recovered by performing the same selection in the \slfit output parameter space.
Follow-up observations and minute modeling allowed a better assessment of the quality of a subset of SL2S candidates that were hence classified with a {\tt confirmed} parameter taking values : 'bad lenses' 0, 'average lenses' 1, 'good lenses' 2 and 'excellent lenses' 3 (G14). For simplicity we merge 'good' and 'excellent' systems into a single class of 'good' lenses. The \RF follow-up allowed the construction of a transfer matrix (G14, Table 4) between the pre-follow-up quality flag {\tt q\_flag} and the {\tt confirmed} parameter. Assuming that SL2S lenses that still lack confirmation (about 400) will obey the same {\tt q\_flag}$\rightarrow${\tt confirmed} transfer matrix we can assign a pseudo-{\tt confirmed} value to all those systems that do not have an actual {\tt confirmed} value.

When applying the cuts defined in \eqref{eq:galcutsID}\footnote{Except that we completely lift here  the requirements on the fit quality ($\chi^2/\nu$) to account for the model limitations at capturing the complex environment of real lens candidates.} and derived from the simulations to the real data, we get 50\% of purity
and only 7\% of completeness, with respect to the parent \RF share between 'good', 'average' and 'bad' lenses. We thus notice a slight improvement in terms of purity as compared to the straight 44\% purity of the \RF sample. On the other hand, the relative completenes is very low. Concerning this result, one has to bear in mind that the selection applied here favors purity over completeness. Moreover, \slfit modeling has a strong prejudice on the possible lens configurations: it is more demanding than \RF and will be less inclusive on real, potentially complex lenses. At this stage it is worse stressing that \slfit modeling is really meant to improve the purity of strong lensing samples in the context of the very large surveys to come.
 
\begin{figure}[h]
\centering
              \includegraphics[width=8.5cm]{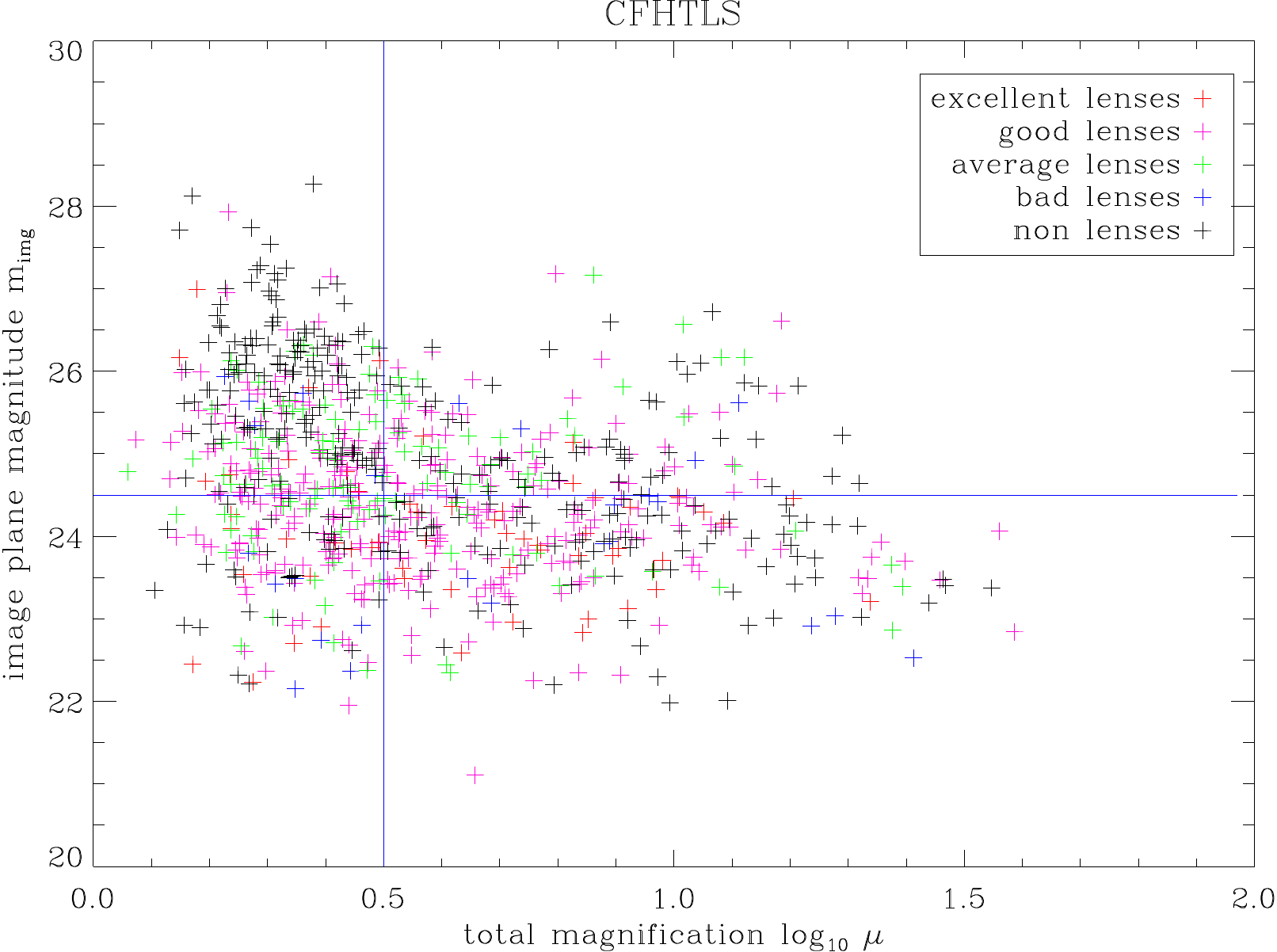}
	   	   \caption{Results of the \slfit modeling of our real lens candidates in the \{$\mu$, $m_{\rm img}$\} subspace with a \galfit subtraction of the deflector. Five classes of lenses are considered : 'excellent lenses' \textit{(red)}, 'good lenses' \textit{(pink)}, 'average lenses' \textit{(green)}, 'bad lenses' \textit{(blue)}, and 'non lenses' \textit{(black)}. This latter population is not part of the final \RF sample and is not considered for calculations. It has been added just to guide the eye to see where 'non lenses' should reside. Lines represent the simplified cuts  that we applied to empirically isolate lenses. They are defined in Eq.~\eqref{eq:cutsIDreal}.}\label{fig:realselec}          
\end{figure}

For now, this modest rate of completeness suggests that the previous selection should be broadened by updating our cuts on the \slfit output parameter space of the real sample. This new selection is substantially simplified to avoid overlaps as much as possible. The distribution of \slfit output parameters for CFHTLS lens candidates is shown in Fig.~\ref{fig:realselec}, where lines trace the following cuts in the \{$\mu$, $m_{\rm img}$\}  subspace:
\begin{equation}\label{eq:cutsIDreal}
\centering
\;\;\;\;\;\;\; \;\;\;\;\;\;\;\;\;\; \{\, \log{(\mu)} > 0.5\,,\;\;\; m_{\rm img} < 24.5\, \}.
\end{equation}

The previous 7\% completeness value is noticeably enhanced since we now achieve a rate of 39\% while leaving the purity rate unchanged around 52\% as reported in Table~\ref{tab:resrobotreal}. Again, we stress that those values are relative to the parent sample preselected with \RF. A more accurate relative distribution between 'good', 'average' and 'bad' lenses should be found in the simulated sample. Indeed, calculations of purity and completeness with the new cuts favoring completeness of Eq.~\eqref{eq:cutsIDreal} give a different picture than for the previous cuts favoring purity of  Eq.~\eqref{eq:galcutsID} and motivated by simulations. We now get 71\% of completeness for mock candidates, but the purity drops to 20\% as also reported in Table~\ref{tab:resrobotreal}. Consequently, we see that a good trade-off between completeness and purity remains difficult to achieve when going through a realistic deflector subtraction as already mentioned in \S~\ref{sec:resmock:galfit}.

\begin{table}[ht]
\centering
\caption{\label{tab:resrobotreal} Updated selection results on real and simulated lens candidates}
\begin{tabular}{lcc}\hline\hline
  &  Real candidates  & Simulations    \\ 
   \hline
  completeness & 39\% & 71\% \\    
\hline
purity & 52\% & 20\% \\    
\hline
 \end{tabular}
\tablefoot{Completeness and purity from the selection defined in Eq.~\eqref{eq:cutsIDreal}, updated on the observed sample, for the 'good' and 'excellent' real lenses combined together \textit{(left)}, and for the simulated 'good lenses' ($\mu_{0} > 4$) in the case of a \galfit subtration  \textit{(right)}.}  
\end{table}

\section{Summary \& Conclusion}\label{sec:conc} 
In this paper we have  attempted to test whether a fast modeling of the light distribution around bright galaxies could provide a valuable diagnostic to decide whether a background source is strongly gravitationally lensed or not.
We have shown that such a modeling and, henceforth, such a diagnostic, are not trivial for ground-based observations, either for simulations of mock strong lensing events or from the CFHTLS survey itself. At present the main limitation resides in the subtraction of the foreground light distribution. However, we have identified the most salient issues and future progress could definitely foster the practical implementation of such a technique to the next generation of surveys. 

With simulations of CFHTLS images in the $g$ band for which the supposed deflector's light is perfectly subtracted before the \slfit analysis is performed, we have shown that:
\begin{itemize}
\item  A fast modeling of  $\sim 10 $ seconds per system can satisfactorily recover the important input parameters, in particular the Einstein Radius $\REin$. This modeling is twice as fast as for the corresponding \hst simulations.
\item Such a modeling allows to precisely isolate lens systems in the \{$\REin$, $\beta$, $\mu$, $m_{\rm img}$, $\chi^{2}/\nu$\} output parameter space of \slfit. Trying to favor purity over completeness at this stage, we estimate that 94\% of purity and 82\% of completeness could be achieved for lensing systems with an Einstein radius large enough, a sufficient magnification, bright lensed images, and a satisfactory fit. Those correspond to Eq.~\eqref{eq:cutsID}.
\end{itemize}

For the more realistic situation in which the foreground light emission is removed by \galfit, the picture is substantially degraded. Focusing on ground-based simulations we found that: 
\begin{itemize}
\item Despite an iterative red-to-blue fitting approach aimed at mitigating the effect of the lensed arcs on the fitting of the foreground galaxy, \galfit often absorbs in its deflector model a large fraction of the lensed light, thus hampering an accurate \slfit fitting of the latter. The median execution time does not change compared to the ideal case.
\item The different classes of lens candidates are much more mixed up in the parameter space, which makes the good lenses much more difficult to isolate from the rest. Therefore, we could only come up with a more heuristic selection as defined by Eq.~\eqref{eq:galcutsID}. It provides samples that are 69\% pure and 31\% complete. We thus notice a substantial loss in terms of selectivity due to our deflector subtraction method implying \galfit.
\end{itemize}

We then applied our modeling and decision strategies to a subset of 517 SL2S lens candidates extracted from the CFHTLS survey :
\begin{itemize}
\item When comparing our fast models to fine models from either \hst or \cfht images that have been previously made by  \cite{PaperIII} for a subset of $\sim$ 45 confirmed lenses, we find a good consistency between the relative estimations of the Einstein radius. This is an encouraging result since the robustness of the $\REin$ estimate is preserved despite the subtraction problems mentionned above and the increased complexity of real lenses that is missing in our simulations (non simple exponential source, non simple de Vaucouleur deflector, neighboring objects along the line of sight, non perfect SIE lens potentials).
\item When filtering this preselected sample through the set of cuts derived from the simulations, we get 50\% of purity -- which is a small gain compared to the 44\% purity for the full original SL2S sample -- and a low rate of completeness of 7\%. We thus attempted to re-adjust the selection on the real data themselves: the simplified set of cuts given by Eq.~\eqref{eq:cutsIDreal} allows to reach 39\% of completeness for a similar purity from the observed sample, but drives the purity down to 20\% (for a completeness of 71\%) when re-applied to the simulations.
\item Predictably, the completeness depends on the lens configurations: our modeling finder preferentially selects lenses with matter and light distributions close to our assumptions (see Sect.\ref{sec:model:assump}). We should thus lose objects for which satellites or extra line-of-sight mass components add much complexity to the lens potential, as well as systems with an intrinsically complex deflector (like a double deflector) or source. Such cases indeed result in a non-trivial image plane flux distribution leading to a chaotic likelihood surface. 
We do not expect the method to be much biased against multiple lenses since two lensed sources will give rise to two independent regions in the parameter space providing a relative gain in $\chi^2$ (or log posterior) despite a poor absolute best-fit $\chi^2$. This is the reason why we did not apply stringent cuts on the fit quality. This essentially amounts to considering the benefit in $\chi^2$ of a model with a lensed source with respect to a model without it, regardless of how close this brings us to the theoretical floor predicted by our noise model. This should however be better quantified with more realistic simulations of this kind of particularly interesting lenses \citep{Gav++08,Son++12,C+A14,Sch14}.
\end{itemize}
Consequently, the massive modeling part of our lens detection method, based on \slfit, is efficient at recovering in a few seconds the most important parameters \{$\REin$, $\mu$, $m_{\rm img}$\}. Because it encodes the lens equation, \slfit modeling incorporates a valuable prior information in the lens finding process \citep[see also][]{Mar++09} and should therefore greatly improve purity (at the expense of complex lenses though).
In its present status, the method is mainly limited by the deflector subtraction, because \galfit assigns flux from the lensed features to the deflector. This hampers an accurate \slfit optimization which, in turn, prevents an effective selection of lenses in the \slfit output parameter space. We foresee a number of options that should significantly improve the process in the near future:
\begin{itemize}
\item We want \slfit to be able to perform a simultaneous fit of the foreground light and of the lensed source in one or more photometric band(s). Coupling the problem of the deflector subtraction with the very lens modeling will allow to fully understand the complex cross-talk at work and, possibly, to reduce some degeneracies. The multi-filter approach has proven valuable in the \RF method (G14).
\item Instead of using a static 0-1 ring-shape mask around each potential deflector to mitigate the effect of neighboring objects on model fitting, we plan to use {\tt SExtractor} \citep{B+A96} to detect and mask out one by one distant sources around the galaxies of interest. Although it appears simple, one should be cautious not to remove lensed features which are not well separated from the central deflector, as it is usually the case. Multiple filters could here also help distinguishing pieces of environment from lensed arcs of similar colors. 
\item Like for the \RF algorithm (G14), a more realistic morphology of foreground deflectors should be included in our simulations since, in real data, we expect a fraction of false positives to arise from edge-on spirals or, more generally, from galaxies exhibiting satellites close to their center. This could be done by superimposing mock arcs on observed non-standard galaxies that are not actually lenses. 
\item The execution time is quite short (of order 10 sec per system) from a modeling perspective, but remains too slow with a view towards lens finding in the context of future large area surveys (all the more so if we go for multi-band simultaneous lens+source model-fitting). A preselection of potential lenses, in the vein of the ETG preselection of the SL2S sample considered here (See Sect.~\ref{ssec:cfhtls}, and G14), will likely remain necessary. We also may have to abandon the benefits of MCMC posterior sampling and improve the straight minimization strategy we briefly mentioned in Sect.~\ref{sssec:optim}. 
\end{itemize}

By simulating the same mock lenses under \cfht and \hst conditions, we were able to estimate the benefit of space-based observations for the prospect of uncovering strong lenses. This is particularly relevant in the context of the future Euclid mission and our \hst simulations are very similar to what is expected with Euclid. This work suggests that the problem of foreground light subtraction is still present (although less prominent) with single band well resolved observations. It should therefore be cautiously addressed for Euclid. Promising methods are being developed \citep{Jos++14} but it remains to be shown whether they will not capture too much flux from the lensed features. Like for ground-based surveys, a multi-band approach could break down the main barriers preventing massive lens modeling with Euclid, and lead to a surge in the population of galaxy-scale strong lenses.


\begin{acknowledgements}
The authors would like to thank the anonymous referee for useful comments and
their colleagues of the SL2S and SLACS collaborations for many useful discussions as well as stimulating exchanges within the Euclid Strong Lensing Science Working Group.
RG also acknowledges E. Bertin for valuable discussions about massive model fitting.
This work was supported by the Centre National des Etudes Spatiales (CNES) and the Programme National Cosmologie et Galaxies (PNCG).
This research is based on observations obtained with MegaPrime/MegaCam, a joint project of CFHT
and CEA/DAPNIA, a joint project of CFHT, Taiwan,
Korea, Canada and France, at the Canada-France-Hawaii Telescope (CFHT) which is operated
by the National Research Council (NRC) of Canada, the Institut National des
Sciences de l'Univers of the Centre National de la Recherche Scientifique
(CNRS) of France, and the University of Hawaii. This work is based in part on
data products produced at TERAPIX and the Canadian Astronomy Data Centre.
\end{acknowledgements}


\bibliographystyle{aa}
\bibliography{references}


\end{document}